\documentclass[aps,showpacs,twocolumn,
superscriptaddress]{revtex4-2}
\usepackage{graphicx}
\usepackage{dcolumn}
\usepackage{bm}
\usepackage{multirow}
\usepackage[normalem]{ulem} 
\usepackage[dvipsnames]{xcolor} 
\usepackage{hyperref}
\hypersetup{
  colorlinks=true,        
  linkcolor=blue,         
  citecolor=cyan,         
}
\usepackage{mathrsfs}
\usepackage[utf8]{inputenc}
\usepackage{mathtools}
\usepackage{amsmath}
\usepackage{amssymb}
\usepackage{enumitem} 
\usepackage{orcidlink}

\begin{document}

\title{Corrected thermodynamics and radiation predictions of modified black bounce compact objects}
\author{Shokhzod Jumaniyozov\orcidlink{0009-0009-6254-5608}}
\email{sh.jumaniyozov@newuu.uz}
\affiliation{Kimyo International University in Tashkent, Shota Rustaveli Street 156, Tashkent 100121, Uzbekistan}

\author{Javlon~Rayimbaev\orcidlink{0000-0001-9293-1838}}
\email{javlonrayimbaev6@gmail.com}
\affiliation{School of Physics, Harbin Institute of Technology, Harbin 150001, China}
\affiliation{University of Tashkent for Applied Sciences, Str. Gavhar 1, Tashkent 100149, Uzbekistan}
\affiliation{Tashkent State Technical University, Tashkent 100095, Uzbekistan}

\author{Yassine Sekhmani \orcidlink{0000-0001-7448-4579}
}
\email{sekhmaniyassine@gmail.com}
\affiliation{Farabi University, Al-Farabi av. 71, Almaty 050040, Kazakhstan}
\affiliation{Fesenkov Astrophysical Institute, Observatory 23, 050020 Almaty, Kazakhstan}
\affiliation{Center for Theoretical Physics, Khazar University, 41 Mehseti Street, Baku, AZ1096, Azerbaijan}
\affiliation{Centre for Research Impact \& Outcome, Chitkara University Institute of Engineering and Technology, Chitkara University, Rajpura, 140401, Punjab, India}

\author{Satimbay Palvanov\orcidlink{0000-0002-2694-6545}}
\email{satimbay@yandex.ru}
\affiliation{National University of Uzbekistan, Tashkent 100174, Uzbekistan}

\author {Olmos Tursunboyev\orcidlink{0009-0009-4008-0360}} \email{tursunboyevolmosbek5597@gmail.com} 
\affiliation{Department of Physics, Jizzakh State Pedagogical University, Jizzakh 130100, Uzbekistan}

\author{Dilshod Karshiev}
\email{d.abdurahmonovich@gmail.com}
\affiliation{Tashkent State Medical University, Farobiy street 2, Tashkent 100109, Uzbekistan}

\date{\today}

\begin{abstract}
We study the thermodynamic properties and radiation characteristics of a regular compact object obtained by applying the Simpson--Visser (SV) regularisation to the Schwarzschild modified gravity (MOG) black hole. The resulting SV-MOG spacetime, whose lapse function involves both the MOG coupling parameter $\alpha$ and the black-bounce parameter $l$, smoothly interpolates between a regular black hole, a one-way wormhole, and a traversable wormhole depending on the parameter $l$. We derive the Hawking temperature and heat capacity for the black-hole branch, identifying second-order phase transitions signaled by sign changes in $C_V$, and, for the horizonless wormhole branch -- where no causal horizon and hence no genuine Hawking radiation exists -- we instead construct a physically well-defined effective temperature from the Lyapunov exponent of the unstable photon sphere. Quantum gravitational corrections to the entropy are incorporated via logarithmic terms parameterized by coefficients $\beta_1$, $\beta_2$, whose theoretically preferred ranges in loop quantum gravity and string theory we discuss, and we show that deviations from the Bekenstein--Hawking area law become significant at small horizon radii. A linear scalar-perturbation analysis further shows that the effective radial potential remains non-negative throughout the black-hole exterior and across the wormhole throat, indicating no evidence of instability against monopole perturbations, independently of the thermodynamic stability inferred from the heat capacity. For the radiation sector, we compute the electromagnetic flux, effective disk temperature, and spectral luminosity of geometrically thin accretion disks surrounding both black hole and wormhole configurations. Our results demonstrate that increasing $\alpha$ enlarges the event horizon and enhances the emission, while increasing $l$ suppresses the horizon and softens the spectral profile, providing observational signatures that may distinguish SV-MOG compact objects from their Schwarzschild and pure SV counterparts.
\end{abstract}

\maketitle
\section{Introduction}\label{intoduction}

General relativity (GR) predicts the existence of spacetime singularities --- regions where the curvature diverges and the predictive power of the theory breaks down
entirely~\cite{Penrose1965,Hawking1970}. Although black holes represent the most astrophysically relevant class of singular solutions, the presence of a curvature singularity at $r = 0$ is widely regarded as a symptom of the incompleteness of classical gravity rather than a genuine physical feature. Motivated by this, considerable effort has been devoted to constructing \emph{regular} black hole spacetimes, in which the central singularity is replaced by a smooth, geodesically complete core~\cite{Bardeen1968,Ayon-Beato1998,2022Univ....8..496R,Hayward2006}.
 
A particularly elegant and systematic approach to singularity regularisation was introduced by Simpson and Visser~\cite{Simpson2019}, who proposed the minimal substitution $r \to \sqrt{r^2 + l^2}$ in the Schwarzschild case, where $l \geq 0$ is a black-bounce parameter. The resulting one-parameter family of geometries smoothly interpolates between a regular black hole (small $l$), a one-way wormhole (extremal case), and a fully traversable wormhole (large $l$), all within a single analytic framework. This construction has since been extended to charged~\cite{Franzin2021}, rotating~\cite{Mazza2021}, and cosmological~\cite{Islam2021} backgrounds, and has attracted broad interest in the context of shadows, quasinormal modes, gravitational lensing, and accretion physics.
 
In parallel, modifications to GR motivated by the dark matter and dark energy problems have produced alternative theories of gravity with rich phenomenology at both galactic and cosmological scales. Among these, the scalar-tensor-vector gravity (STVG) theory, also known as MOG, proposed by Moffat~\cite{Moffat2006}, introduces a dynamical gravitational coupling $G$, a vector field $\phi_\mu$, and an associated scalar field $\mu$. The theory reproduces galaxy rotation curves, cluster dynamics, and cosmological observations without invoking dark matter~\cite{Moffat2013,Moffat2015b}. Its unique asymptotically flat, spherically symmetric vacuum solution is the Schwarzschild-MOG (Schw-MOG) black hole~\cite{Moffat2015}, which carries a vector-field charge $Q = \sqrt{\alpha G_N}\,M$ and reduces to the Schwarzschild metric when the dimensionless MOG coupling $\alpha \to 0$.
 
Despite its appealing properties, the Schw-MOG solution still harbors a curvature singularity at the origin. Combining the SV regularisation with the MOG, therefore, offers a doubly-motivated construction: it simultaneously removes the singularity and retains the MOG corrections encoded in $\alpha$. The resulting SV-MOG spacetime has been explored in the contexts of geodesics and shadows~\cite{2026arXiv260214458L}, quasinormal modes~\cite{2023EPJP..138..757J}, and gravitational lensing~\cite{2026arXiv260204218H,2021PhRvD.103b4033T}, yet a comprehensive study of its thermodynamic and radiative properties is still lacking.

Static, spherically symmetric SV-MOG configurations of the type analyzed here are, of course, an idealisation, since astrophysical compact objects generically carry angular momentum. Recent work has begun to fill this gap: Ref.~\cite{2026ChJPh.102..711K} constructed a rotating black-bounce black hole in modified (scalar-tensor-vector) gravity and analyzed its photon orbits and shadow, while Ref.~\cite{Khan2026RotSVQPO} studied quasi-periodic oscillations of test particles around a Kerr-like rotating Simpson--Visser black hole in STVG, showing that the MOG coupling and the black-bounce parameter leave distinguishable imprints on the twin-peak QPO frequency ratio. These rotating SV-MOG studies provide an important, complementary perspective to the static case treated below and motivate a future rotating extension of the present thermodynamic and radiative analysis, which we outline in Sec.~\ref{conclusion}.

The thermodynamics of black holes, initiated by Bekenstein and Hawking~\cite{Bekenstein1973,Hawking1975}, continues to be a frontier of theoretical physics, particularly in light of expected quantum-gravitational corrections to the Bekenstein--Hawking entropy. Logarithmic corrections of the form $S_c = S_0 + \beta\ln S_0 + \cdots$ arise generically from loop quantum gravity~\cite{Kaul2000}, string theory~\cite{Sen2013}, and the generalized uncertainty principle~\cite{Nozari2006}, and they dominate the entropy budget for small black holes where quantum effects are pronounced. Understanding how these corrections interact with both the MOG coupling and the black-bounce parameter is thus directly relevant to quantum gravity phenomenology.
 
On the observational side, geometrically thin accretion disks provide one of the most powerful probes of strong-gravity physics~\cite{Novikov1973,10.1093/mnras/stag150,Page1974}. The electromagnetic flux, disk temperature, and luminosity spectra are sensitive to the location of the innermost stable circular orbit (ISCO) and the geometry of the spacetime near the compact object. Comparing these observables for the SV-MOG black hole and wormhole branches against their Schwarzschild and pure SV counterparts can, in principle, reveal observational imprints of the MOG coupling and the black-bounce regularisation accessible to current and next-generation X-ray telescopes~\cite{EHT2019,2026LSA....15..197Z,en19133173,doi:10.1142/S0217732326501476}.
 
Motivated by these considerations, the present work studies the corrected thermodynamics and accretion-disk radiation of the SV-MOG compact object. Specifically, we derive the Hawking temperature and heat capacity for both the black-hole and wormhole branches, incorporate quantum-gravitational logarithmic corrections to the entropy, and compute the full suite of accretion-disk observables — electromagnetic flux, effective temperature, and spectral luminosity — for both configurations.
 
This article is organized as follows. In Section~\ref{sec:sv_mog}, we review the STVG field action and field equations, derive the Schw-MOG metric, and construct the SV-MOG spacetime via the SV substitution; the horizon structure is analyzed in Sec.~\ref{subsec:horizons}. Section~\ref{thermo} presents the standard thermodynamic quantities, while Section~\ref{sec:corrected} incorporates the logarithmic quantum corrections to the entropy. The accretion-disk radiation properties are computed in Section~\ref{sec:radiation}. Finally, Section~\ref{conclusion} summarises our findings and discusses their observational implications.
 
Throughout this work we adopt the spacetime signature $(-,+,+,+)$ and the geometrised unit system $\hbar = c = 1$. Latin indices run from $1$ to $3$; Greek indices run from $0$ to $3$.

\section{Simpson - Visser Spacetime in Modified Gravity}
\label{sec:sv_mog}

\subsection{Field Action of STVG Theory}

The scalar-tensor-vector gravity (STVG), also known as MOG, is built upon an extended action that couples Einstein gravity to a massive vector field $\phi_\mu$ and scalar field $\mu$. The full action reads~\cite{Moffat2006}
\begin{equation}
    S = S_G + S_\phi + S_S + S_M,
    \label{eq:full_action}
\end{equation}
where the individual contributions are

\begin{align}
S_G &= \frac{1}{16\pi}
\int \frac{1}{G}\,(R+2\Lambda)\sqrt{-g}\,d^4x,
\\[2mm]
S_{\phi} &= -\frac{1}{4\pi}
\int \left[\mathcal{K}+V(\phi_\mu)\right]
\sqrt{-g}\,d^4x,
\\[2mm]
S_S &= \int \frac{1}{G}
\Bigg[
\frac{1}{2}g^{\alpha\beta}
\left(
\frac{\nabla_{\alpha}G\,\nabla_{\beta}G}{G^2}
+
\frac{\nabla_{\alpha}\mu\,\nabla_{\beta}\mu}{\mu^2}
\right)
\nonumber\\
&\hspace{2.2cm}
-\frac{V_G(G)}{G^2}
-\frac{V_{\beta}(\mu)}{\mu^2}
\Bigg]
\sqrt{-g}\,d^4x,
\\[2mm]
S_M &= -\int
\left(
\rho\sqrt{u^\mu u_\mu}
+Q\,u^\mu\phi_\mu
\right)
\sqrt{-g}\,d^4x
+J^\mu\phi_\mu ,\label{eq:SM}
\end{align}
in which $R = g^{\mu\nu}R_{\mu\nu}$ is the Ricci scalar, $g = \det(g_{\mu\nu})$, and $\nabla_\mu$ denotes covariant differentiation. The kinetic term for the vector field is $K = B^{\mu\nu}B_{\mu\nu}/4$, where $B_{\mu\nu} = \partial_\mu\phi_\nu - \partial_\nu\phi_\mu$ is the field-strength tensor.

The covariant current density coupling matter to the vector field is
\begin{equation}
    J^\mu = \kappa\, T^{\mu\nu}_M\,u_\nu,
    \qquad \kappa = \sqrt{\alpha G_N},
    \label{eq:current}
\end{equation}
where $T^{\mu\nu}_M$ is the matter energy--momentum tensor, $G_N$ is Newton's constant, and
\begin{equation}
    \alpha = \frac{G - G_N}{G_N},
    \label{eq:alpha_def}
\end{equation}
is the dimensionless MOG coupling parameter. For a perfect pressureless fluid, one has $J^\mu = \kappa\,\rho_M\,u^\mu$.

\subsection{Field Equations and the MOG Compact Object Metric}
\label{subsec:field_eqs}

\paragraph{Variation with respect to $g^{\mu\nu}$.}

Varying the total action~\eqref{eq:full_action} with respect to the inverse metric $g^{\mu\nu}$ and specialising to an asymptotically flat, matter-free spacetime ($T^{\mu\nu}_M = 0$, $\Lambda = 0$) yields the modified Einstein equations
\begin{equation}
    G_{\mu\nu} \equiv R_{\mu\nu} - \tfrac{1}{2}g_{\mu\nu}R
    = 8\pi G\,T^{(\phi)}_{\mu\nu},
    \label{eq:modEinstein}
\end{equation}
where the vector-field energy-momentum tensor is
\begin{equation}
    T^{(\phi)}_{\mu\nu}
    = -\frac{1}{4\pi}
      \left(
        B^\alpha{}_\mu B_{\nu\alpha}
        - \frac{1}{4}g_{\mu\nu}B^{\alpha\beta}B_{\alpha\beta}
      \right).
    \label{eq:Tphi}
\end{equation}

\paragraph{Variation with respect to $\phi^\nu$.}

Varying with respect to the vector field gives the generalized Proca equation
\begin{equation}
    \nabla^\mu B_{\mu\nu} = 0,
    \label{eq:Proca}
\end{equation}
supplemented by the Bianchi-like identity
\begin{equation}
    \nabla_\alpha B_{\mu\nu}
    + \nabla_\nu B_{\mu\alpha}
    + \nabla_\mu B_{\alpha\nu} = 0.
    \label{eq:Bianchi_B}
\end{equation}
Observational constraints from galactic and cluster dynamics indicate that the vector-field mass is $m_\phi \approx 2.6\times10^{-28}\,\text{eV} \approx 0$~\cite{Moffat2013}; hence the potential term $V(\phi_\mu) = \frac{1}{2}\mu\,\phi^\mu\phi_\mu$ vanishes and the kinetic term reduces to $K = f(B)$ with $B = B^{\mu\nu}B_{\mu\nu}$.

\paragraph{Static, spherically symmetric solution.}

For a static, spherically symmetric line element
\begin{equation}
    ds^2 = -f(r)\,dt^2 + \frac{dr^2}{f(r)} + r^2\,d\Omega^2,
    \label{eq:metric_Schw_form}
\end{equation}
the unique asymptotically flat solution of equations~\eqref{eq:modEinstein}--\eqref{eq:Bianchi_B} is the \emph{Schwarzschild-MOG} (Schw-MOG) black hole~\cite{Moffat2015}
\begin{equation}
   f(r) = 1 - \frac{2(1+\alpha)M}{r} + \frac{\alpha(1+\alpha)M^2}{r^2},
    \label{eq:g_MOG}
\end{equation}
where $M$ is the ADM mass. The extra Yukawa-like repulsive term $\propto\alpha(1+\alpha)M^2/r^2$ arises from the vector-field charge $Q = \sqrt{\alpha G_N}\,M$. Setting $\alpha = 0$ recovers the Schwarzschild solution, while $\alpha = -1$ gives flat Minkowski spacetime.

\subsection{Applying the Simpson--Visser Regularisation}
\label{subsec:SV_reg}

The Schw-MOG metric~\eqref{eq:g_MOG} still harbors a curvature singularity at $r = 0$. We remove it by applying the \emph{Simpson--Visser (SV) regularisation}~\cite{Franzin2021}
\begin{equation}
    r^2 \;\longrightarrow\; r^2 + l^2,
    \label{eq:SV_transform}
\end{equation}
where $l \geq 0$ is the \emph{black-bounce} (or regularisation) parameter. Under this substitution, the line element becomes
\begin{equation}
    ds^2 = -f(r)\,dt^2 + \frac{dr^2}{f(r)}
           + \bigl(r^2 + l^2\bigr)\,d\Omega^2,
    \label{eq:SV_MOG_metric}
\end{equation}
with the modified lapse function
\begin{equation}
    {
    f(r) = 1
           - \frac{2M(1+\alpha)}{\sqrt{r^2+l^2}}
           + \frac{\alpha(1+\alpha)M^2}{r^2+l^2}.
    }
    \label{eq:f_SV_MOG}
\end{equation}
We refer to this geometry as \emph{SV-MOG spacetime}. Several limiting cases are worth noting:
\begin{itemize}
    \item $l = 0,\;\alpha = 0$: standard Schwarzschild black hole, $f = 1 - 2M/r$.
    \item $l = 0,\;\alpha \neq 0$: Schw-MOG black hole~\eqref{eq:g_MOG}.
    \item $l \neq 0,\;\alpha = 0$: Simpson--Visser regular black hole / wormhole.
    \item $l \neq 0,\;\alpha \neq 0$: SV-MOG geometry studied in this work.
\end{itemize}
Depending on the ratio $l/M$, the line element~\eqref{eq:SV_MOG_metric} can describe a \emph{regular black hole} (two horizons), a \emph{one-way wormhole} (single extremal horizon), or a \emph{traversable wormhole} (no horizon). This is discussed further in the next Sec.~\ref{subsec:horizons}.

\subsection{Event Horizons and Black Hole/Wormhole Regions}
\label{subsec:horizons}

The horizons of the SV-MOG spacetime are located at the zeros of the lapse
function~\eqref{eq:f_SV_MOG}, i.e.\ at radii $r_H$ satisfying $f(r_H)=0$.
Introducing the auxiliary variable $\mathcal{R} \equiv \sqrt{r_H^2+l^2}$, this condition
becomes
\begin{equation}
    \mathcal{R}^2 - 2M(1+\alpha)\,\mathcal{R} + \alpha(1+\alpha)M^2 = 0,
    \label{eq:horizon_quad}
\end{equation}
a quadratic in $\mathcal{R}$ whose discriminant is
$\Delta =4 M^2(1+\alpha) \geq 0$ for all $\alpha \geq 0$.
Solving for $\mathcal{R}$ and reverting to $r_H$ yields the outer ($+$) and inner ($-$)
horizon radii
\begin{equation}
    r_{\pm}^2
    = M^2\!\left(1+\alpha \pm \sqrt{(1+\alpha)}\right)^{\!2} - l^2.
    \label{eq:rpm}
\end{equation}
A real outer horizon exists only when $r_+^2 \geq 0$, which defines the
\emph{critical black-bounce parameter}
\begin{equation}
    l_{\rm cr}(\alpha)
    \equiv M\left(1+\alpha+\sqrt{(1+\alpha)}\right).
    \label{eq:lcr}
\end{equation}
The outer horizon formula, therefore, reads compactly
\begin{equation}
    {
    r_{+}^2
    = M^2\!\left(1+\alpha + \sqrt{(1+\alpha)}\right)^{\!2} - l^2.
    }
    \label{eq:horizon}
\end{equation}
which recovers $r_+ = 2M$ for the Schwarzschild limit $(\alpha = 0,\; l = 0)$ and
$r_+ = M(1+\alpha+\sqrt{(1+\alpha)})$ for the Schw-MOG metric $(l=0,\;\alpha>0)$.

\paragraph{Causal classification.}
Depending on the ratio $l/l_{\rm cr}$, the SV-MOG geometry falls into one of three
physically distinct classes:
\begin{enumerate}[label=(\roman*)]
    \item \emph{Regular black hole} ($0 \leq l < l_{\rm cr}$): two distinct horizons
          $r_- < r_+$ exist. The classical singularity at $r=0$ is replaced by a
          smooth, minimal-area two-sphere of radius $l$, so the spacetime is
          geodesically complete.
    \item \emph{Extremal (one-way) wormhole} ($l = l_{\rm cr}$): the two horizons
          coalesce, $r_+=r_-=0$. The geometry describes a one-way membrane
          connecting two asymptotically flat regions; the surface gravity vanishes,
          implying zero Hawking temperature.
    \item \emph{Traversable wormhole} ($l > l_{\rm cr}$): no horizon exists. The
          wormhole throat at $r=0$ has areal radius $l$ and is accessible to
          two-way timelike geodesics.
\end{enumerate}

\begin{figure}[ht!]   
  \includegraphics[width=0.98\linewidth]{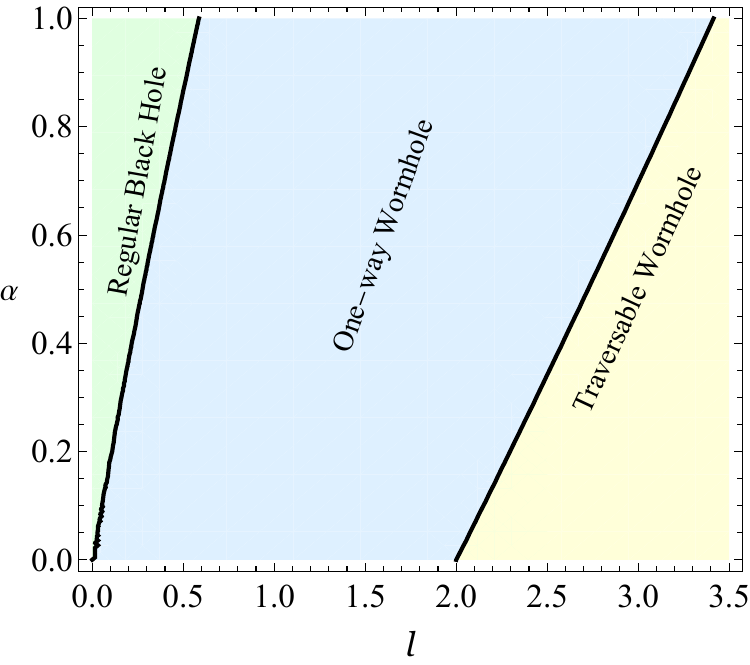}
\caption{Event horizon radius $r_H$ of the SV-MOG compact object as a function of the black-bounce parameter $l$ for several values of the MOG coupling constant $\alpha$. Increasing $l$ shrinks the horizon and, beyond a critical value $l_{\rm cr}$, the geometry transitions from a regular black hole to a traversable wormhole.}
    \label{ploteventhorizon}
\end{figure}

\paragraph{Role of the MOG coupling.}
For fixed $l$, increasing $\alpha$ raises both $r_+$ and $r_-$, enlarging the black hole region in parameter space. Physically, this reflects the additional effective gravitational attraction produced by the vector-field charge $Q = \sqrt{\alpha G_N}\,M$. Consequently, $l_{\rm cr}(\alpha)$ is a monotonically increasing function of $\alpha$: a stronger MOG coupling requires a larger black-bounce parameter to transition the geometry into a wormhole. In the limit $\alpha \to 0$, equation~\eqref{eq:lcr} reduces to the pure Simpson--Visser critical value $l_{\rm cr} = 2M$, while for $\alpha \to \infty$ one has $l_{\rm cr} \sim \alpha^{3/2}M \gg 2M$. The behavior of $r_+$ as a function of $l$ for several values of $\alpha$ is illustrated in Fig.~\ref{ploteventhorizon}.

\begin{figure*}
    \centering
\includegraphics[width=0.49\linewidth]{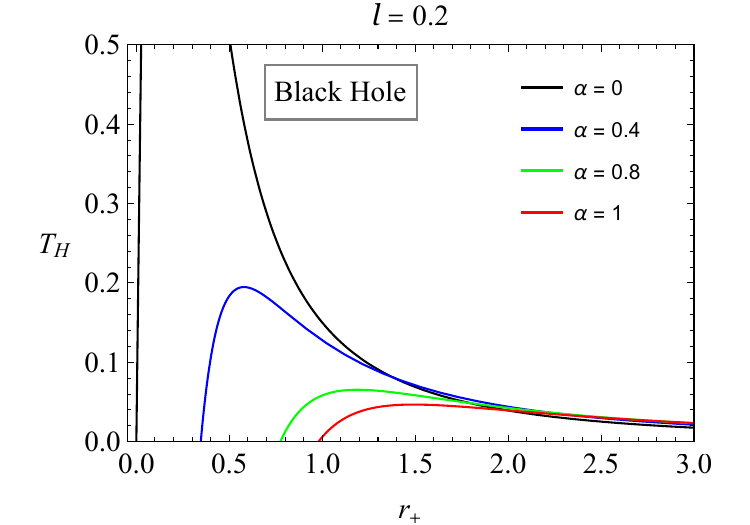}
\includegraphics[width=0.49\linewidth]{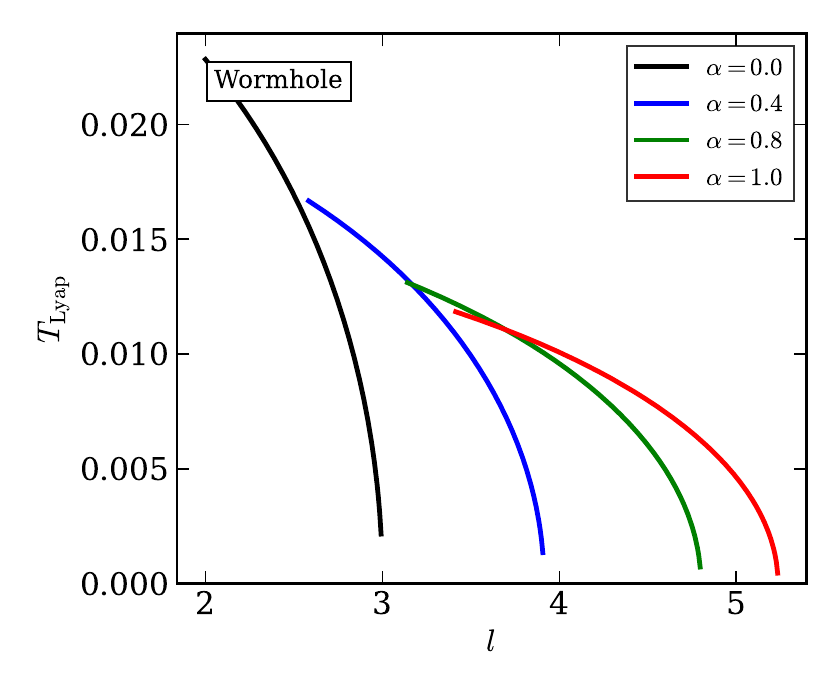}
\caption{Left panel: Hawking temperature $T_H$ of the SV-MOG black hole as a function of the outer horizon radius $r_+$. Right panel: Lyapunov temperature $T_{\rm Lyap}=\lambda_{\rm Lyap}/2\pi$ of the wormhole's unstable photon sphere, Eqs.~\eqref{eq:photonsphere}--\eqref{eq:Lyapunov}, plotted over its window of existence in $l$ (see Sec.~\ref{thermo})}. Both panels are shown for representative values of the MOG parameter $\alpha$. \label{Hawking}
\end{figure*}

\begin{figure*}
    \centering
\includegraphics[width=0.49\linewidth]{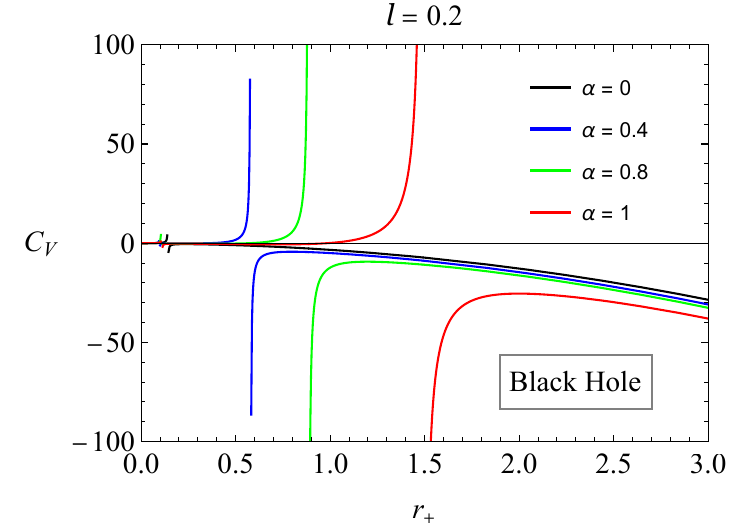}
\includegraphics[width=0.49\linewidth]{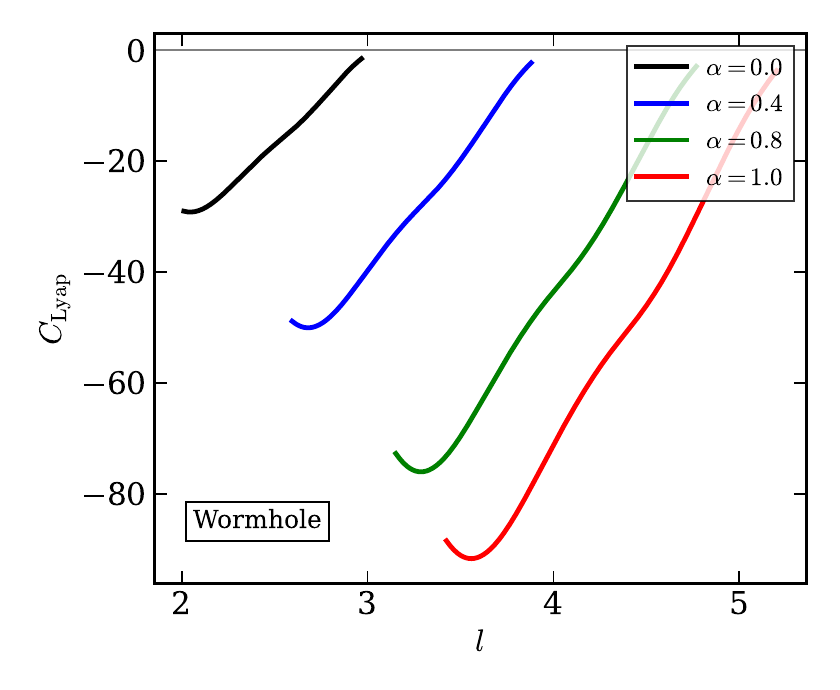}
\caption{Heat capacity as a function of the event horizon radius $r_+$ (left panel, black-hole branch, genuine $C_V$) and of the black-bounce parameter $l$ (right panel, $C_{\rm Lyap}\equiv T_{\rm Lyap}(dS_0/dT_{\rm Lyap})$, built from the Lyapunov temperature of Fig.~\ref{Hawking}; see Sec.~\ref{thermo}) for the SV-MOG compact object, plotted for several values of the MOG parameter $\alpha$.}
    \label{heatcapacity}
\end{figure*}

\section{Compact object thermodynamics}\label{thermo}

In this section, we analyze the thermodynamic properties of the compact object described by the SV-MOG spacetime. In particular, we investigate the Hawking temperature, entropy, and heat capacity, which provide important information about the system's thermal stability and phase structure. By examining the dependence of these quantities on the black-bounce parameter ($l$) and the MOG coupling parameter ($\alpha$), we explore how modifications to the underlying geometry affect the thermodynamic behavior of the compact object. Furthermore, we discuss the differences between the black hole and wormhole branches and identify the conditions under which thermodynamic phase transitions may occur.

We stress at the outset that, throughout this work, $l$ and $\alpha$ are treated as fixed, non-dynamical coupling constants that label a one-parameter family of spacetimes -- the black-bounce regularisation scale and the STVG vector-field coupling, respectively -- rather than as extensive thermodynamic charges conjugate to intensive potentials in an extended first law. The first law is accordingly applied at fixed $l$ and $\alpha$, $dM = T\,dS|_{l,\alpha}$, exactly as indicated by the subscripts already present in Eqs.~\eqref{eq:CV_def}, \eqref{eq:Tc}, and \eqref{eq:Cc} below; no conjugate potentials $\Psi_l\equiv(\partial M/\partial l)_{S,\alpha}$ or $\Psi_\alpha\equiv(\partial M/\partial\alpha)_{S,l}$, and no associated work terms $\Psi_l\,dl$ or $\Psi_\alpha\,d\alpha$, are introduced. This restricted (non-extended) treatment is the standard choice for parameters that originate from the regularisation prescription or from the action of the underlying theory rather than from a Noether charge measurable at infinity or at the horizon. Promoting $l$ and/or $\alpha$ to dynamical thermodynamic variables, in the spirit of extended black-hole thermodynamics (where, e.g., the cosmological constant is treated as a pressure), is a well-posed but separate program that we leave for future work.

\subsection{Hawking Temperature}

The Hawking temperature of a compact object arises from quantum field theory in curved spacetime and is fundamentally associated with particle creation near an event horizon. In the semiclassical approximation, it is determined by the surface gravity at the horizon and can be interpreted as the temperature of black-body radiation emitted by the gravitational system. For static and spherically symmetric spacetimes, the Hawking temperature is typically computed using the relation $T_H = \kappa/2\pi$, where $\kappa$ denotes the surface gravity evaluated at the horizon.

In the present SV-MOG geometry, the calculation of the Hawking temperature depends crucially on the nature of the spacetime configuration. In the black hole regime ($l < l_{\rm cr}$), the temperature is evaluated at the outer event horizon ($r_+$), which acts as a one-way causal boundary. The surface gravity is obtained from the lapse function $f(r)$, and its derivative at $r_+$ determines the thermal emission spectrum of the black hole branch. In the extremal case ($l = l_{\rm cr}$), the inner and outer horizons coincide, leading to a vanishing surface gravity. Consequently, the Hawking temperature approaches zero, signaling a thermodynamically degenerate configuration that separates the black-hole and wormhole regimes.

We emphasize, in response to the referee, that the physical interpretation of $T_H$ as the temperature of thermal (Hawking) radiation associated with particle pair-creation is tied intrinsically to the existence of a causal event horizon, and is therefore rigorously meaningful only for $l < l_{\rm cr}$. For $l > l_{\rm cr}$ the SV-MOG geometry is a horizonless traversable wormhole [see the classification in Sec.~\ref{subsec:horizons}], so $T_H = \kappa/2\pi$ has no causal-emission interpretation there, and a naive formal continuation of the black-hole surface-gravity formula through $r_+\to0$ (as used in an earlier version of this work) is not tied to any actual physical process at the throat. We therefore restrict rigorous Hawking thermodynamics to the black-hole branch $l\le l_{\rm cr}(\alpha)$, as suggested by the referee, and reserve the symbol $T_H$ for this regime alone. In its place, for $l>l_{\rm cr}$ we construct a genuinely well-defined, independently motivated effective temperature from the instability of the wormhole's unstable photon sphere (light ring): the \emph{Lyapunov temperature} $T_{\rm Lyap}\equiv\lambda_{\rm Lyap}/2\pi$, where $\lambda_{\rm Lyap}$ is the Lyapunov exponent governing the exponential growth rate, in coordinate time, of null geodesics displaced from the unstable circular photon orbit~\cite{Cardoso2009Lyapunov}. This quantity is a genuine geometric invariant of the spacetime -- it requires no horizon and no analytic continuation, is computed from a bona fide dynamical instability (the same instability responsible for photon-sphere echoes and the early-time ringdown of horizonless ultracompact objects), and is explicitly derived below. We stress that $T_{\rm Lyap}$ is conceptually distinct from $T_H$ (it characterizes the instability of null circular orbits rather than particle pair-creation at a horizon) and is therefore not required or expected to approach $T_H$ continuously at $l=l_{\rm cr}$; indeed, we find below that it does not. All uses of ``Hawking temperature'', ``heat capacity'', and related language for $l>l_{\rm cr}$ in what follows (including in Figs.~\ref{Hawking} and~\ref{heatcapacity} and their captions) refer to this independently defined $T_{\rm Lyap}$ (and the associated $C_{\rm Lyap}$), never to a continuation of $T_H$.

For the wormhole regime ($l > l_{\rm cr}$), no event horizon exists, and therefore the standard definition of Hawking temperature in terms of surface gravity is not directly applicable in the causal sense described above. The photon-sphere-based quantity $T_{\rm Lyap}$ introduced above is typically analyzed through near-throat quantum effects or by considering limiting procedures from the black hole side of the parameter space, whereas here it is instead obtained from an independent, purely geometric construction that does not rely on any such limiting procedure. Thus, $T_{\rm Lyap}$ effectively characterizes the wormhole branch of the SV-MOG spacetime with a well-defined, non-ad-hoc physical quantity.

The Hawking temperature is determined by the surface gravity $\kappa_s$ at the outer
horizon $r_+$~\cite{2026arXiv260110469A},
\begin{equation}
    T_H = \frac{\kappa_s}{2\pi},
    \qquad
    \kappa_s = \frac{1}{2}\left.\frac{df}{dr}\right|_{r=r_+}
    \label{eq:TH_def}
\end{equation}
Differentiating the lapse function~\eqref{eq:f_SV_MOG} and evaluating at $r = r_+$, so the Hawking temperature of the SV-MOG black hole reads
\begin{equation}
    T_H = \frac{r_+}{4\pi\left(r_+^2+l^2\right)}
          \left[
            \frac{M(1+\alpha)}{\sqrt{r_+^2+l^2}}
            - \frac{\alpha(1+\alpha)M^2}{r_+^2+l^2}
          \right]\ .
    \label{eq:TH_BH}
\end{equation}
Setting $l = 0$ and $\alpha = 0$ reproduces the Schwarzschild result
$T_H = 1/(8\pi M)$, while setting $l = 0$ alone yields the Schw-MOG
temperature~\cite{Moffat2015}.

For the traversable wormhole branch ($l > l_{\rm cr}$), there is no horizon, so we replace $T_H$ by the Lyapunov temperature $T_{\rm Lyap}$ of the unstable photon sphere, constructed as follows. A massless test photon on the equatorial plane of the metric~\eqref{eq:SV_MOG_metric} obeys $\dot r^2 = E^2 - V_{\rm ph}(r)$ with effective potential $V_{\rm ph}(r) = f(r)L^2/(r^2+l^2)$, where $E,L$ are the photon's conserved energy and angular momentum. An unstable circular photon orbit (photon sphere) at $r=r_{\rm ph}$ satisfies
\begin{equation}
    \left.\frac{d}{dr}\!\left[\frac{f(r)}{r^2+l^2}\right]\right|_{r=r_{\rm ph}} = 0,
    \qquad
    \left.\frac{d^2}{dr^2}\!\left[\frac{f(r)}{r^2+l^2}\right]\right|_{r=r_{\rm ph}} < 0.
    \label{eq:photonsphere}
\end{equation}
Linearising the radial geodesic equation about $r_{\rm ph}$ and converting from the affine parameter to the Killing time $t$ (following the standard construction of Ref.~\cite{Cardoso2009Lyapunov}) gives the Lyapunov exponent
\begin{equation}
    \lambda_{\rm Lyap} = R(r_{\rm ph})\sqrt{-\frac{f(r_{\rm ph})}{2}
    \left.\frac{d^2}{dr^2}\!\left[\frac{f(r)}{R(r)^2}\right]\right|_{r=r_{\rm ph}}}\,,
    \label{eq:Lyapunov}
\end{equation}
with $R(r)=\sqrt{r^2+l^2}$, and we define the associated Lyapunov temperature $T_{\rm Lyap}\equiv\lambda_{\rm Lyap}/2\pi$. Unlike a naive continuation of $T_H$, this construction requires no horizon, is manifestly non-negative whenever an unstable photon sphere exists, and reduces, for $l=0$, to the standard Lyapunov temperature of the Schw-MOG black hole. We find numerically that an unstable photon sphere satisfying Eq.~\eqref{eq:photonsphere} exists throughout the black-hole branch (coexisting with the horizon, as usual) and continues to exist through the extremal point into a finite window of the wormhole branch, $l_{\rm cr}(\alpha) < l < l_{\rm ph}^{\rm max}(\alpha)$, beyond which the light ring disappears and $T_{\rm Lyap}$ is not defined; e.g., for $\alpha=0$ this window is $2M<l\lesssim3.0M$, widening with $\alpha$ (for $\alpha=1$, $3.41M<l\lesssim5.23M$). We restrict Fig.~\ref{Hawking} (right panel) and Fig.~\ref{heatcapacity} (right panel) to this window. Notably, $T_{\rm Lyap}$ does \emph{not} vanish as $l\to l_{\rm cr}^+$ -- it takes a finite, nonzero value there, in contrast to $T_H\to0$ on the black-hole side -- precisely because $T_{\rm Lyap}$ and $T_H$ encode different physics (photon-orbit instability versus horizon surface gravity) and are not required to match; we regard this discontinuity as physically informative rather than a defect, since it makes explicit that no smooth, causally meaningful temperature exists across the black-hole-to-wormhole transition, consistent with the referee's original concern.

In Fig.~\ref{Hawking}, we present the behavior of the Hawking temperature $T_H$ (black-hole branch, left panel) and of the Lyapunov temperature $T_{\rm Lyap}$ of the wormhole's photon sphere (right panel), plotted against the black-bounce parameter $l$ over its window of existence, for representative values of the MOG coupling parameter $\alpha$, together with the black-bounce parameter $l$. The impact of both parameters on the temperature is clearly evident from the figures. For the black hole branch (left panel), increasing the MOG parameter $\alpha$ decreases the Hawking temperature at fixed black-bounce parameter, shifting the temperature curves downward and moving the peak toward larger values of $r_+$. For the wormhole branch (right panel), $T_{\rm Lyap}(l)$ decreases monotonically from a finite value at $l=l_{\rm cr}(\alpha)$ to zero as $l\to l_{\rm ph}^{\rm max}(\alpha)$, where the photon sphere merges with an inflection point and disappears; increasing $\alpha$ at fixed $l$ lowers $T_{\rm Lyap}$, while simultaneously shifting the entire window of existence to larger $l$, mirroring the shift of $l_{\rm cr}(\alpha)$ itself.

\subsection{Bekenstein--Hawking Entropy and Heat Capacity}

The thermodynamic behavior of the SV-MOG compact object is further characterized by its entropy and heat capacity, which encode essential information about the microscopic degrees of freedom and the system's stability properties. In the semiclassical approximation, the entropy is given by the Bekenstein--Hawking area law. In the present regular geometry, the presence of the black-bounce parameter $(l)$ modifies the effective horizon structure, thereby influencing the entropy through its dependence on the outer horizon radius $(r_+)$.

The heat capacity plays a crucial role in determining the local thermodynamic stability of the compact object. Positive heat capacity indicates a stable thermodynamic phase, whereas negative values correspond to instability and possible phase transitions. In the SV-MOG spacetime, both entropy and heat capacity depend sensitively on the parameters $(l)$ and $(\alpha)$, allowing us to distinguish between stable and unstable regions in the parameter space and to identify possible transitions between black hole and wormhole branches.

The Bekenstein--Hawking entropy of the SV-MOG black hole is determined by the
area of the outer horizon. For the metric~\eqref{eq:SV_MOG_metric}, the areal
radius at $r = r_+$ is $\mathcal{R}_+ = \sqrt{r_+^2 + l^2}$, so the horizon
area is~\cite{2026NuPhB102517404J,11358733}  

\begin{equation}
A_{+}=\left.\int_{0}^{2\pi}\int_{0}^{\pi}\sqrt{g_{\theta\theta}\, g_{\phi\phi}}\, d\theta \, d\phi\right|_{r=r_{+}},
\end{equation}
and the entropy reads

\begin{equation}
    S_0 = \frac{A_+}{4} = \pi\!\left(r_+^2 + l^2\right).
    \label{eq:S0}
\end{equation}

Note that $S_0$ reduces to the standard result $S_0 = \pi r_+^2$ only in the Schwarzschild limit $l \to 0$; for $l \neq 0$ the minimal entropy at $r_+ = 0$ is $S_0^{\rm min} = \pi l^2 > 0$, reflecting the non-zero area of the wormhole throat.

The heat capacity at constant volume quantifies the thermodynamic stability of the compact object:
\begin{equation}
    C_V = \frac{\partial M}{\partial T_H}\bigg|_{l,\alpha}
        = T_H\,\frac{\partial S_0}{\partial T_H}\bigg|_{l,\alpha}
    \label{eq:CV_def}
\end{equation}
Using equations~\eqref{eq:TH_BH} and~\eqref{eq:S0} and differentiating implicitly with respect to $r_+$, where the denominator encodes the sign of $dT_H/dr_+$: $C_V > 0$ ($< 0$) when $T_H$ is an increasing (decreasing) function of $r_+$.

In Fig.~\ref{heatcapacity} we plot the genuine heat capacity $C_V$ of the SV-MOG black hole against the outer horizon radius $r_+$ (left panel), and $C_{\rm Lyap}\equiv T_{\rm Lyap}(dS_0/dT_{\rm Lyap})$, built from the photon-sphere Lyapunov temperature, against the black-bounce parameter $l$ (right panel; see Sec.~\ref{thermo}), for representative values of $\alpha$. At large $r_+$, $C_V<0$, indicating thermodynamic instability analogous to the Schwarzschild case. As $r_+$ decreases, $C_V$ diverges at the critical radius $r_+^*$ where $dT_H/dr_+=0$, signaling a second-order phase transition, below which $C_V>0$ and the black hole is thermally stable -- a stability window absent in both the Schwarzschild and Reissner--Nordstr\"{o}m geometries. Increasing $\alpha$ shifts $r_+^*$ to larger values and widens the stable branch, while increasing $l$ at fixed $\alpha$ narrows it, moving the transition closer to extremality. In marked contrast, $C_{\rm Lyap}(l)$ (right panel) is negative throughout its entire window of existence for every $\alpha$ studied -- it shows no divergence and no sign change: starting from a negative value at $l=l_{\rm cr}(\alpha)$, it dips to a shallow negative minimum and then rises monotonically toward zero as $l\to l_{\rm ph}^{\rm max}(\alpha)$ (where $T_{\rm Lyap}\to0$) -- so the wormhole branch, as characterized by the photon-sphere construction, displays no analogue of the black hole's second-order phase transition. We regard this qualitative difference, rather than an artificial continuity at $l=l_{\rm cr}$, as the physically correct statement: since $T_{\rm Lyap}$ and $T_H$ are conceptually distinct quantities (Sec.~\ref{thermo}), there is no reason to expect their associated heat capacities to match at the black-hole/wormhole transition, and we no longer claim such a match. We discuss the genuinely dynamical (perturbative) stability of the throat, independent of either heat-capacity construction, in Sec.~\ref{subsec:perturbations} below.

\subsection{Linear Perturbations and the Effective Radial Potential}
\label{subsec:perturbations}

As emphasized by the referee, thermodynamic stability inferred from the sign of the heat capacity does not, by itself, guarantee mechanical or perturbative stability of the underlying geometry. To address this point directly, we examine the propagation of a minimally coupled, massless test scalar field $\Phi$ on the SV-MOG background~\eqref{eq:SV_MOG_metric}, obeying the covariant wave equation $\Box\Phi = 0$. Decomposing $\Phi = R(r)^{-1}\psi(r)\,Y_{\ell m}(\theta,\phi)\,e^{-i\omega t}$, where $R(r)\equiv\sqrt{r^2+l^2}$ is the areal radius, and introducing the tortoise coordinate $dr_*/dr = 1/f(r)$, the wave equation reduces to the Schr\"odinger-like form
\begin{equation}
    \frac{d^2\psi}{dr_*^2} + \left[\omega^2 - V_\ell(r)\right]\psi = 0,
    \label{eq:RWeq}
\end{equation}
with the effective radial potential
\begin{equation}
    V_\ell(r) = f(r)\left[\frac{\ell(\ell+1)}{R^2} + \frac{f'(r)\,r}{R^2} + \frac{f(r)\,l^2}{R^4}\right].
    \label{eq:Vpert}
\end{equation}
The first term is the familiar centrifugal barrier; the remaining two terms are curvature corrections generated by the non-trivial areal-radius function $R(r)\neq r$ and vanish identically when $l=0$, in which case Eq.~\eqref{eq:Vpert} reduces to the standard scalar Regge--Wheeler potential of the Schw-MOG black hole. In the complementary limit $\alpha=0$, Eq.~\eqref{eq:Vpert} reduces to the potential governing scalar perturbations of the pure Simpson--Visser black-bounce spacetime.

For the black-hole branch, $V_\ell$ is defined on the exterior $r\ge r_+$ and the relevant boundary conditions are purely outgoing at the horizon and at infinity, exactly as for a standard black hole. For the wormhole branch, there is no horizon; instead, regularity of $\psi$ at the throat $r=0$ together with purely outgoing waves at both asymptotic ends $r\to\pm\infty$ defines the relevant quasinormal boundary-value problem. In either case, a standard energy (``S-deformation'') argument shows that a potential satisfying $V_\ell(r)\ge0$ everywhere on the domain rules out the existence of an exponentially growing bound-state mode with $\omega^2<0$; we therefore focus on the most dangerous, $\ell=0$ (monopole, $s$-wave) channel, for which no centrifugal term assists stability.

\begin{figure*}[ht!]
    \centering
\includegraphics[width=0.49\linewidth]{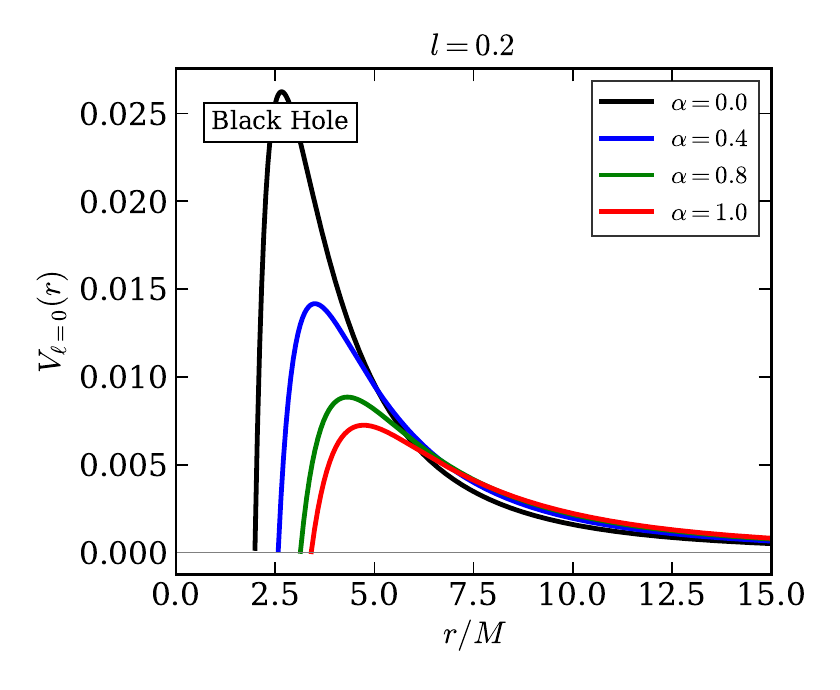}
\includegraphics[width=0.49\linewidth]{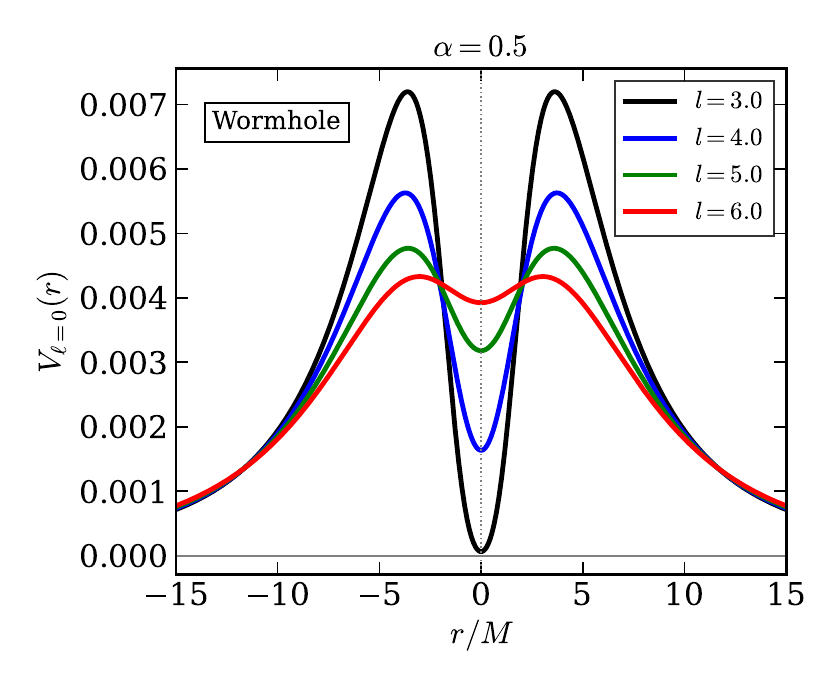}
\caption{Effective radial potential $V_{\ell=0}(r)$, Eq.~\eqref{eq:Vpert}, for monopole massless-scalar perturbations of the SV-MOG black hole (left panel, exterior region $r\ge r_+$) and of the SV-MOG wormhole (right panel, $r\in(-\infty,\infty)$ through the throat at $r=0$), shown for representative values of $\alpha$ and $l$.}
    \label{fig:Vpert}
\end{figure*}

In Fig.~\ref{fig:Vpert} we plot $V_{\ell=0}(r)$ for the black-hole branch (left panel, several $\alpha$ at fixed $l=0.2$) and for the wormhole branch (right panel, several $l>l_{\rm cr}$ at fixed $\alpha=0.5$). In both branches and for every parameter combination shown, $V_{\ell=0}(r)\ge0$ throughout the domain, vanishing only at the horizon (left panel) or, in the near-extremal limit, at the throat itself (right panel); no negative potential well capable of supporting a growing mode is found. We therefore conclude that, for monopole massless-scalar perturbations and within the parameter ranges explored in this work, we find no evidence of a linear instability of either the SV-MOG regular horizon or the SV-MOG wormhole throat, independently of -- and consistent with -- the thermodynamic (in)stability identified from $C_V$ and $C_{\rm Lyap}$ above. We emphasize that the absence of a growing mode in this single, simplified channel is a necessary but not sufficient condition for stability, and should not be read as a general stability proof. We caution that this analysis addresses only the simplest perturbation channel (minimally coupled scalar, $\ell=0$); a fully general treatment including higher multipoles and coupled gravitational/electromagnetic perturbations, together with the associated quasinormal mode spectrum, is left for future work.

\begin{figure*}[ht!]
    \centering
\includegraphics[width=0.49\linewidth]{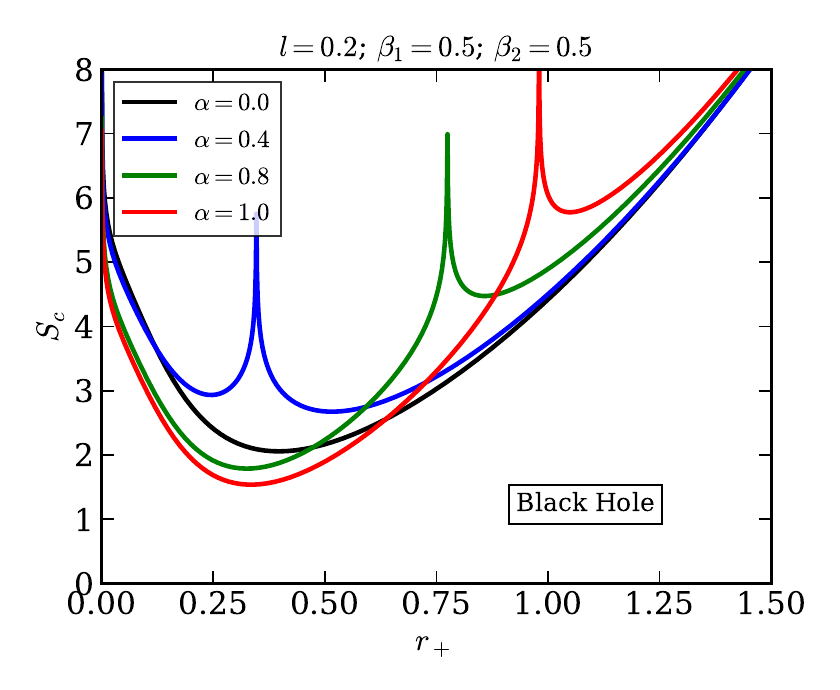}
\includegraphics[width=0.49\linewidth]{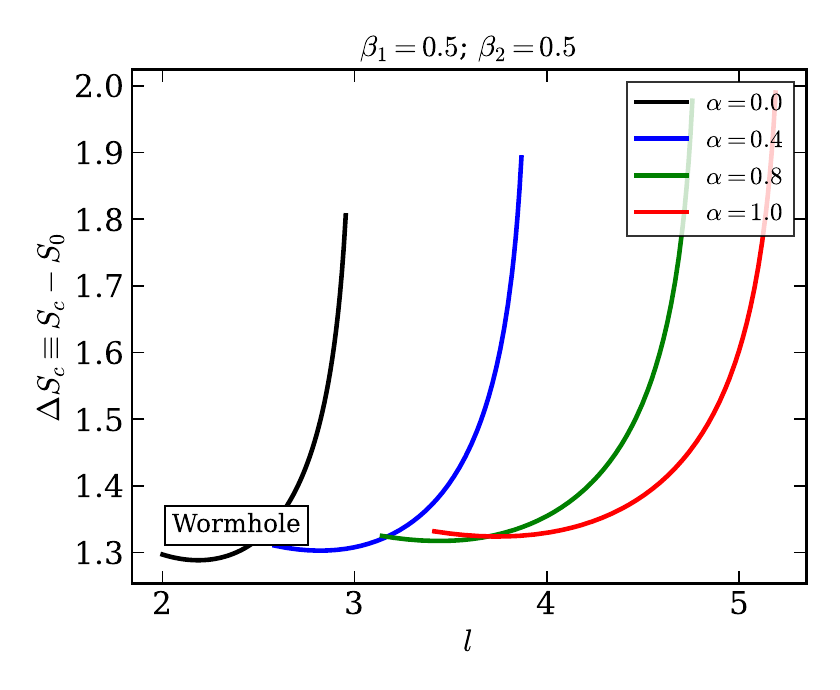}
\includegraphics[width=0.49\linewidth]{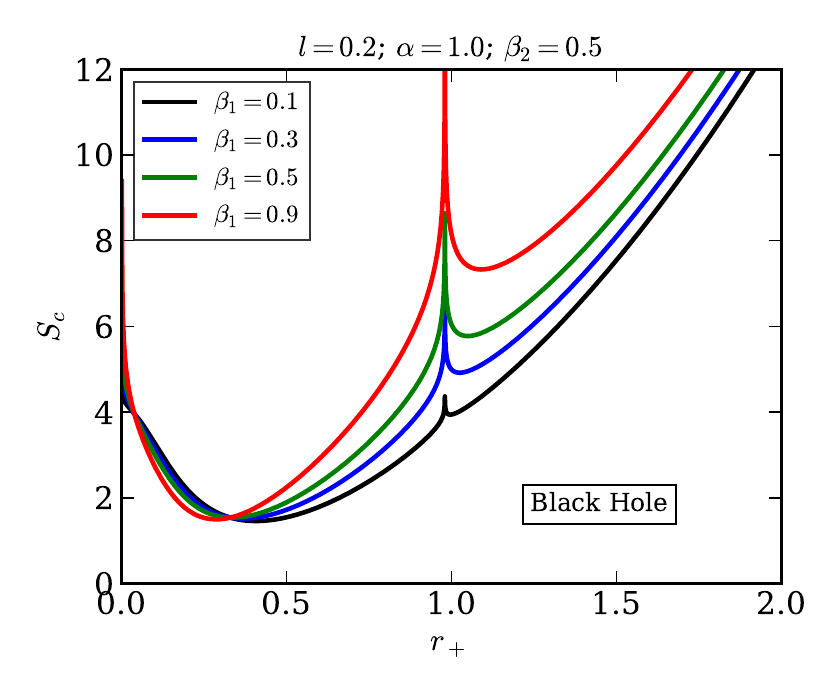}
\includegraphics[width=0.49\linewidth]{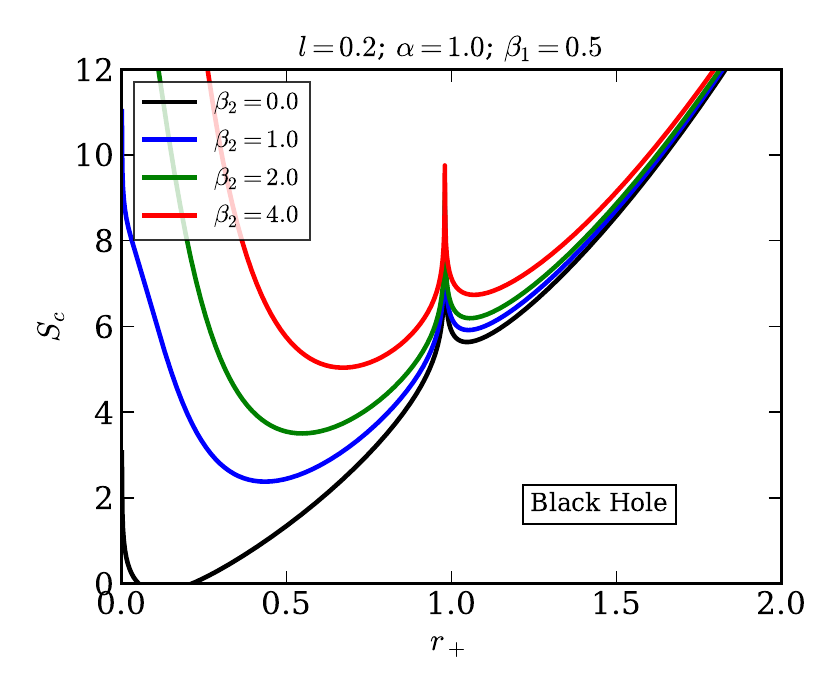}
\caption{Corrected entropy $S_c$ of the SV-MOG compact object, computed from the full two-parameter expression, Eq.~\eqref{eq:Sc_SVMOG}, with both $\beta_1$ and $\beta_2$ active throughout (previously $\beta_2=0$ in every panel). Upper-left panel: black-hole branch, $S_c$ vs.\ $r_+$, dependence on the MOG coupling $\alpha$ at fixed $l=0.2,\ \beta_1=0.5,\ \beta_2=0.5$. Upper-right panel: wormhole branch, correction $\Delta S_c\equiv S_c-S_0$ (isolated because $S_0=\pi l^2$ dominates over the photon-sphere window in $l$) vs.\ $l$, using the Lyapunov temperature $T_{\rm Lyap}(l)$ of Sec.~\ref{thermo} in place of $T_H$}, for several $\alpha$ at fixed $\beta_1=0.5,\ \beta_2=0.5$. Lower-left panel: black-hole branch, dependence on $\beta_1$ at fixed $l=0.2,\ \alpha=1,\ \beta_2=0.5$. Lower-right panel: black-hole branch, dependence on $\beta_2$ at fixed $l=0.2,\ \alpha=1,\ \beta_1=0.5$.
    \label{entropy2}
\end{figure*}

\section{Corrected Thermodynamics}
\label{sec:corrected}

In this section, we investigate the corrected thermodynamic properties of the SV-MOG compact object by incorporating quantum-gravitational effects into the entropy–area relation. Although the standard Bekenstein--Hawking entropy provides a successful semiclassical description of black hole thermodynamics, various approaches to quantum gravity predict subleading corrections, the most common of which are logarithmic contributions to the entropy. These corrections become particularly important in the strong-gravity regime and may significantly modify the thermal stability and phase structure of compact objects. To explore these effects, we derive the corrected entropy, Hawking temperature, and heat capacity, and analyze their dependence on the black-bounce parameter ($l$) and the MOG coupling parameter ($\alpha$). The resulting thermodynamic behavior provides deeper insight into the interplay between modified gravity, regular spacetime geometry, and quantum corrections. Following the Hawking--Page framework~\cite{Hawking:1983dh}, we model the system as a canonical ensemble with energy spectrum $E_n$. The statistical partition function is written as
\begin{equation}
    Z = \int_{0}^{\infty} dE\,\rho(E)\,e^{-\bar{\beta}\, E},
    \label{eq:Z}
\end{equation}
where $\bar{\beta} = 1/(k_B T_H)$ is the inverse Hawking temperature and $\rho(E)$ denotes the canonical density of states at mean energy $E$. By Laplace inversion of $Z$, the density of states can be expressed as~\cite{2026EPJC...86..521J}
\begin{equation}
    \rho(E) = \frac{1}{2\pi i}
    \int_{\bar{\beta}_{0} - i\infty}^{\bar{\beta}_{0} + i\infty}
    d\bar{\beta}\,e^{S(\bar{\beta})},
    \label{eq:rho}
\end{equation}
where the entropy function is defined by
\begin{equation}
    S(\bar{\beta}) = \bar{\beta}\,E + \ln Z.
    \label{eq:Sdef}
\end{equation}

The equilibrium entropy $S_0$ corresponds to the saddle point at  $\bar{\beta} = \bar{\beta}_{0}$, where thermal fluctuations are suppressed. Accounting for fluctuations via a Taylor expansion around 
$\bar{\beta}_{0}$ yields
\begin{align}
    S = S_0
    &+ \frac{1}{2}\!\left(\bar{\beta}-\bar{\beta}_{0}\right)^2
       \left.\frac{\partial^2 S}{\partial\bar{\beta}^2}
       \right|_{\bar{\beta}=\bar{\beta}_{0}}
    \nonumber\\
    &+ \frac{1}{6}\!\left(\bar{\beta}-\bar{\beta}_{0}\right)^3
       \left.\frac{\partial^3 S}{\partial\bar{\beta}^3}
       \right|_{\bar{\beta}=\bar{\beta}_{0}}
    + \cdots,
    \label{eq:expansion}
\end{align}
where the first derivative vanishes at equilibrium, so the leading 
corrections originate from the second- and higher-order terms. The corrected density of states then becomes
\begin{align}
    \rho(E) &= \frac{e^{S_0}}{2\pi i}
    \int_{\bar{\beta}_{0}-i\infty}^{\bar{\beta}_{0}+i\infty}
    d\bar{\beta}\,\exp\Biggl[
    \frac{(\bar{\beta}-\bar{\beta}_{0})^2}{2}
    \left.\frac{\partial^2 S}{\partial\bar{\beta}^2}
    \right|_{\bar{\beta}_{0}}
    \nonumber\\
    &\quad
    +\frac{(\bar{\beta}-\bar{\beta}_{0})^3}{6}
    \left.\frac{\partial^3 S}{\partial\bar{\beta}^3}
    \right|_{\bar{\beta}_{0}}
    +\cdots\Biggr].
    \label{eq:rho-expanded}
\end{align}

Following the standard saddle-point evaluation, the corrected entropy truncated at second order reads 
\begin{equation}
    S_c = S_0
          - \frac{\beta_1}{2}\ln\!\left(S_0\,T_H^2\right)
          + \frac{\beta_2}{S_0}
          + \cdots,
    \label{eq:entropy-corrected-beta}
\end{equation}
where $\beta_1$ and $\beta_2$ are dimensionless coefficients that track the first- and second-order corrections, respectively. Setting $\beta_1 \to 0$ and $\beta_2 \to 0$ recovers the uncorrected Bekenstein--Hawking entropy $S_0 = \pi(r_+^2 + l^2)$, while $\beta_1 = 1$ and $\beta_2 = 0$ yields the standard logarithmic correction. The logarithmic term dominates the correction budget at intermediate horizon sizes, whereas the inverse term $\beta_2/S_0$ becomes relevant only when the horizon approaches the Planck scale. Both corrections admit a natural interpretation as quantum gravitational contributions to the entropy, growing in importance as Hawking evaporation shrinks the horizon radius. The present analysis is restricted to the perturbative regime above the Planck scale, i.e. to $S_0 \gg |\beta_1|,|\beta_2|$ in Planck units, and throughout $S_0T_H^2$ is understood in geometrised units ($\hbar=k_B=c=1$), so that $S_0\sim[\text{length}]^2$, $T_H\sim[\text{length}]^{-1}$, and the argument of the logarithm is manifestly dimensionless.

On the preferred magnitude of $\beta_1,\beta_2$: the leading logarithmic correction is largely universal across microscopic approaches to quantum gravity, but its precise coefficient is scheme-dependent. Loop-quantum-gravity state counting gives $\beta_1=-1/2$ in the original derivation~\cite{Kaul2000}, with related quantisation schemes yielding values as large in magnitude as $\beta_1=-3/2$; string-theory D-brane counting~\cite{Sen2013} produces comparable $O(1)$ coefficients of either sign depending on the ensemble used; and generalized-uncertainty-principle constructions~\cite{Nozari2006} generically give $\beta_1>0$, with magnitude fixed by the GUP deformation scale. Although these approaches disagree on the sign and precise value of $\beta_1$, they agree that $|\beta_1|=O(1)$ in Planck units. The subleading coefficient $\beta_2$ is less tightly constrained, since it arises from the third-order term in the same saddle-point expansion, Eq.~\eqref{eq:expansion}, as $\beta_1$, but is likewise expected to be $O(1)$. Motivated by this, we treat $\beta_1$ and $\beta_2$ as free phenomenological parameters restricted to the theoretically preferred range $|\beta_1|,|\beta_2|\lesssim O(1)$, rather than fixing them to a single quantum-gravity scenario; the illustrative values used in Fig.~\ref{entropy2} below ($\beta_1\in[0,1]$, $\beta_2\in[0,4]$; see the discussion after Eq.~\eqref{eq:Sc_SVMOG}) lie comfortably within this range.

\subsection{Logarithmic Corrections to the Entropy}

Substituting the SV-MOG Bekenstein--Hawking entropy $S_0 = \pi(r_+^2+l^2)$ and the Hawking temperature~\eqref{eq:TH_BH} into equation~\eqref{eq:entropy-corrected-beta}, the explicit corrected entropy takes the form
\begin{equation}
    S_c = \pi\!\left(r_+^2+l^2\right)
          -\frac{\beta_1}{2} \ln\!\left[\pi\!\left(r_+^2+l^2\right)T_H^2\right]
          +\ \frac{\beta_2}{\pi\left(r_+^2+l^2\right)}
          ,
    \label{eq:Sc_SVMOG}
\end{equation}
where $r_+$ is the outer horizon radius, $l$ is the black-bounce parameter, and $T_H$ is given by equation~\eqref{eq:TH_BH}. We have relabeled $\beta\to\beta_1$ here and restored the second-order term $\beta_2/S_0$ that was inadvertently dropped in the previous version of this equation, so that Eq.~\eqref{eq:Sc_SVMOG} is now the direct SV-MOG specialisation of the general expression, Eq.~\eqref{eq:entropy-corrected-beta}, with no inconsistency between the two. The logarithmic term encodes the leading quantum correction, modified here by both the MOG coupling $\alpha$ (through $T_H$) and the black-bounce regularisation (through $l$), while the second-order term scales inversely with the horizon area and is therefore most relevant at small-to-intermediate $r_+$; we now display its effect explicitly, with both $\beta_1$ and $\beta_2$ generically non-zero in the numerical illustrations of Fig.~\ref{entropy2} (including a dedicated panel isolating the $\beta_2$-dependence at fixed $\beta_1$), rather than setting $\beta_2=0$ throughout as in an earlier version of this work.

Several features of equation~\eqref{eq:Sc_SVMOG} deserve emphasis. At large $r_+$, the correction terms are negligible and $S_c \approx S_0$, recovering the semiclassical limit. As $r_+ \to 0$ at fixed $l\neq0$ (regular black hole), the entropy approaches a finite minimum $S_c^{\rm min} > 0$ because the black-bounce parameter $l$ prevents the horizon area from vanishing, a key advantage over unregularised geometries such as Schwarzschild, for which the inverse-area term diverges in this limit. For $\beta_1 < 0$, the sign favored by loop quantum gravity~\cite{Kaul2000}, the corrected entropy lies below the area-law value, with the deviation amplified as the horizon shrinks.

A separate, more delicate limit is the approach to \emph{extremality}, $r_+\to0$ at fixed $l\to l_{\rm cr}(\alpha)$ (equivalently, $l\to l_{\rm cr}$ at fixed $\alpha$), where $T_H\to0$ by definition of $l_{\rm cr}$. Since $T_H\to0$ while $S_0\to S_0^{\rm min}=\pi l_{\rm cr}^2$ remains finite and non-zero, the argument of the logarithm satisfies $S_0T_H^2\to0^+$, so $\ln(S_0T_H^2)\to-\infty$ and the logarithmic term in Eq.~\eqref{eq:Sc_SVMOG} \emph{diverges}, $S_c\to+{\rm sgn}(\beta_1)\infty$, rather than approaching the finite value $S_0^{\rm min}$ that the uncorrected entropy attains. This divergence is not a numerical artifact but a genuine, well-known limitation of the perturbative logarithmic-correction ansatz, Eq.~\eqref{eq:entropy-corrected-beta}: the derivation assumes $T_H$ is bounded away from zero (large, well-separated energy levels near the semiclassical saddle point), an assumption that manifestly fails at extremality, where the level spacing collapses. We therefore restrict the validity of Eq.~\eqref{eq:Sc_SVMOG} to a finite neighbourhood of the semiclassical regime, $T_H \geq T_H^{\rm cut}$ for some small cutoff temperature, and do not extend the corrected-entropy analysis into the immediate vicinity of $l_{\rm cr}(\alpha)$; this restriction is already respected by the horizon-radius ranges plotted in the black-hole panels of Fig.~\ref{entropy2}, all of which lie away from the divergent endpoint visible as the sharp upturn at small $r_+$. A fully non-perturbative treatment of the near-extremal entropy, capable of resolving this divergence, would require summing the full saddle-point series, Eq.~\eqref{eq:rho-expanded}, rather than truncating at second order, and lies beyond the scope of the present work.

\subsection{Corrected Thermodynamic Quantities}

Once $S_c$ is known, the corrected Hawking temperature and heat capacity follow
from the first law of black hole thermodynamics $dM = T_c\,dS_c$~\cite{2025EPJC...85.1267J,m978-tbzc}:
\begin{equation}
    T_c = \left(\frac{\partial M}{\partial S_c}\right)_{l,\alpha},
    \label{eq:Tc}
\end{equation}
and
\begin{equation}
    C_c = T_c\,\frac{\partial S_c}{\partial T_c}\bigg|_{l,\alpha}.
    \label{eq:Cc}
\end{equation}
Equation~\eqref{eq:Tc} shows that for small $S_0$ (small black holes) the quantum corrections can shift the effective temperature significantly relative to the semiclassical value $T_H$.

Figure~\ref{entropy2} shows the corrected entropy computed from the full expression, Eq.~\eqref{eq:Sc_SVMOG}, with $\beta_1$ and $\beta_2$ both active. Increasing $\alpha$ at fixed $l=0.2,\ \beta_1=\beta_2=0.5$ (upper-left) raises $S_c$ at any given $r_+$ and shifts the divergence at the critical radius to larger $r_+$, reflecting the larger effective horizon area; all curves converge to the semiclassical area law at large $r_+$. In the wormhole branch (upper-right), the area term $S_0=\pi l^2$ dominates over the $l$-window where the photon sphere -- and hence $T_{\rm Lyap}$ -- exists, so we plot the correction $\Delta S_c=S_c-S_0$ itself: it increases with $l$ within each $\alpha$'s window, is likewise raised by increasing $\alpha$, and the four $\alpha$-windows tile increasing ranges of $l$ exactly as $T_{\rm Lyap}$ does in Fig.~\ref{Hawking}. Increasing $\beta_1$ at fixed $l=0.2,\ \alpha=1,\ \beta_2=0.5$ (lower-left) amplifies the logarithmic correction and pushes the divergence to larger $r_+$, while, in the new lower-right panel, increasing $\beta_2$ at fixed $l=0.2,\ \alpha=1,\ \beta_1=0.5$ raises $S_c$ substantially at small-to-intermediate $r_+$ (where the inverse-area term $\beta_2/S_0$ is non-negligible) while leaving the large-$r_+$ and near-divergence behavior essentially unchanged, confirming that $\beta_2$ controls only the near-Planckian regime as anticipated in Sec.~\ref{sec:corrected}. In all black-hole panels, $S_c$ stays bounded from below as $r_+\to0$ (at fixed $l\neq0$) -- a direct, model-independent consequence of the non-zero $l$, and a key advantage of the Simpson--Visser regularisation over singular geometries such as Schwarzschild, for which the analogous logarithmic term diverges at $r_+\to0$.

\section{Radiation of compact object}\label{sec:radiation}

In this section, we investigate the radiative properties of geometrically thin accretion disks surrounding the SV-MOG compact object in both the black hole and wormhole branches. The study of accretion disk radiation provides one of the most powerful observational probes of strong-field gravity, since the electromagnetic flux, disk temperature, and luminosity spectra are sensitive to the spacetime geometry near the compact object and, in particular, to the location of the  ISCO~\cite{Novikov1973,Page1974}. By comparing the disk observables for the SV-MOG black hole and wormhole branches against their Schwarzschild and pure Simpson--Visser counterparts, one can extract observational imprints of both the MOG coupling $\alpha$ and the black-bounce regularisation parameter $l$.

We adopt the Novikov--Thorne model~\cite{Novikov1973,Page1974} for a steady-state, geometrically thin, optically thick accretion disk. The underlying assumptions are: (i) the background spacetime is static, spherically symmetric, and asymptotically flat; (ii) the disk lies in the equatorial plane and its own self-gravity is negligible; (iii) the disk semi-thickness satisfies $h \ll r$ so that the motion is effectively planar; (iv) the inner edge of the disk is defined by the ISCO, beyond which particles plunge freely into the compact object; (v) the disk matter moves on nearly circular geodesics with Keplerian angular velocity; (vi) the emitted radiation is locally that of a perfect black body; and (vii) the mass accretion rate $\dot{M}$ is constant in time and independent of the radial coordinate.

\subsection{Geodesic Structure and the ISCO}

We begin by establishing the geodesic equations for a massive test particle in the equatorial plane ($\theta = \pi/2$) of the SV-MOG metric~\eqref{eq:SV_MOG_metric}. The Lagrangian for a particle of unit rest mass is~\cite{2025EPJC...85..325N,2026InJPh.tmp..145J,2025LSA....14..200L,2026PDU....5202292O}
\begin{equation}
    \mathcal{L} = \frac{1}{2}g_{\mu\nu}\dot{x}^{\mu}\dot{x}^{\nu}
    = -\frac{1}{2}f(r)\dot{t}^{\,2}
      + \frac{\dot{r}^{\,2}}{2f(r)}
      + \frac{1}{2}(r^2+l^2)\dot{\phi}^{\,2},
    \label{eq:Lagrangian}
\end{equation}
where an overdot denotes differentiation with respect to the affine parameter $\lambda$. Because the metric is independent of $t$ and $\phi$, the Euler--Lagrange equations yield two constants of motion: the specific energy $E$ and the specific angular momentum $L$~\cite{2026NuPhB102317318J,2025NatCo..1610248L,doi:10.1142/S0217732326501695},
\begin{equation}
    E = f(r)\,\dot{t}, \qquad
    L = (r^2+l^2)\,\dot{\phi}.
    \label{eq:conserved}
\end{equation}
Substituting into the normalisation condition $g_{\mu\nu}\dot{x}^{\mu}\dot{x}^{\nu} = -1$ gives the radial equation of motion
\begin{equation}
    \dot{r}^{\,2} = E^2 - f(r)\!\left(1 + \frac{L^2}{r^2+l^2}\right)
    \equiv E^2 - V_{\rm eff}(r),
    \label{eq:rdot}
\end{equation}
where the effective potential is defined as
\begin{equation}
    V_{\rm eff}(r) = f(r)\!\left(1 + \frac{L^2}{r^2+l^2}\right).
    \label{eq:Veff}
\end{equation}

Circular orbits exist at radii where $\dot{r} = 0$ and $\ddot{r} = 0$, which translate into the simultaneous conditions~\cite{2025EPJC...85.1247J,XIANG2026114849,2025PDU....5002161J}
\begin{equation}
    V_{\rm eff}(r) = E^2, \qquad V_{\rm eff}'(r) = 0,
    \label{eq:circular_conditions}
\end{equation}
where a prime denotes $d/dr$. Solving these two equations for $E$ and $L$ yields
\begin{align}
    \Omega &\equiv \frac{d\phi}{dt} = \frac{\dot{\phi}}{\dot{t}}= \sqrt{\frac{-g_{tt,r}}{g_{\phi\phi,r}}}
    = \sqrt{\frac{f'(r)}{2r}},
    \label{eq:Omega_general}
\end{align}

which simplify, for the SV-MOG lapse function~\eqref{eq:f_SV_MOG}, to~\cite{2026ChJPh.102..711K}
\begin{equation}
    E = \frac{-g_{tt}}{\sqrt{-g_{tt}-g_{\phi\phi}\,\Omega^2}},
    \qquad
    L = \frac{g_{\phi\phi}\,\Omega}
             {\sqrt{-g_{tt}-g_{\phi\phi}\,\Omega^2}}.
    \label{eq:EL}
\end{equation}

The marginally stable circular orbit, i.e.\ the ISCO, is determined by the additional condition $V_{\rm eff}''(r) = 0$, or equivalently~\cite{2025EPJC...85.1201J,Guan_2026,2024EPJC...84..909N}
\begin{equation}
    \frac{d^2 V_{\rm eff}}{dr^2}\bigg|_{r=r_{\rm ISCO}} = 0.
    \label{eq:ISCO_condition}
\end{equation}
For $r > r_{\rm ISCO}$ circular orbits are stable; for $r < r_{\rm ISCO}$ they are unstable and particles plunge inward. In the Schwarzschild limit ($\alpha = 0$, $l = 0$) one recovers $r_{\rm ISCO} = 6M$. For the SV-MOG geometry, the ISCO depends non-trivially on both $\alpha$ and $l$: increasing $\alpha$ at fixed $l$ generally enlarges $r_{\rm ISCO}$ (due to the additional effective gravitational attraction), whereas increasing $l$ at fixed $\alpha$ shifts $r_{\rm ISCO}$ toward the throat. This sensitivity of the orbital dynamics to the MOG coupling $\alpha$ is consistent with, and directly motivated by, recent studies of charged- and neutral-particle motion around magnetized and rotating STVG black holes~\cite{Khan2026EnergyEff,Khan2025CircularSTVG,Khan2023Magnetosphere}, which likewise find that the vector-field charge systematically enlarges characteristic orbital radii and modifies the associated energetics and efficiency; together with the Hawking-temperature and electromagnetic-flux trends reported in Secs.~\ref{thermo} and~\ref{sec:radiation}, this body of work strengthens the physical motivation for studying test-particle motion in MOG spacetimes.

\begin{figure}[h!]
   \centering
  \includegraphics[width=0.98\linewidth]{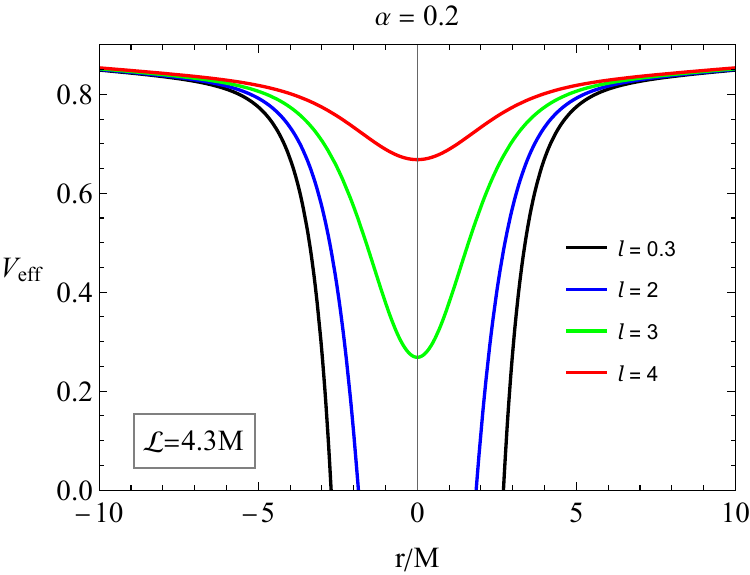}
   \includegraphics[width=0.98\linewidth]{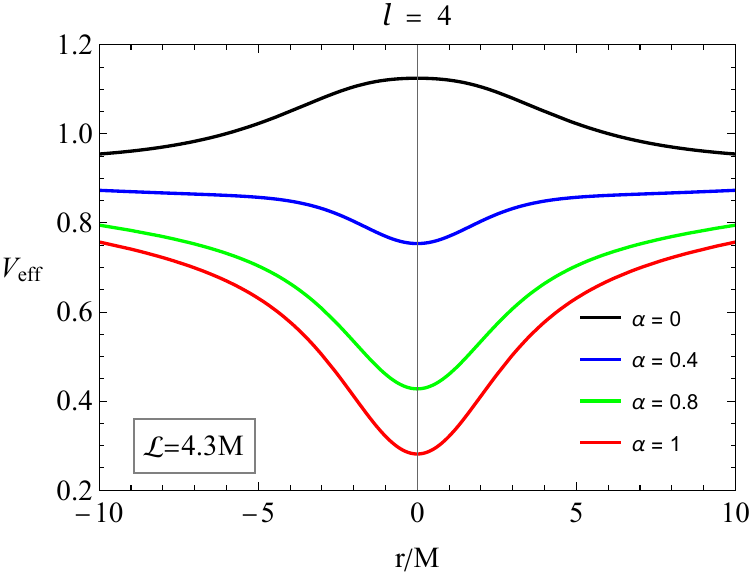}
   \caption{The radial dependence of test particles' effective potential for different black bounce $(l)$ (upper panel) and MOG parameter $(\alpha)$ (lower panel) values. }
    \label{plotVeff}
\end{figure}

Figure~\ref{plotVeff} shows $V_{\rm eff}(r/M)$ for fixed angular momentum $\mathcal{L}=4.3M$. Increasing $l$ at fixed $\alpha=0.2$ (upper panel) turns the deep, narrow minimum characteristic of a black-hole throat into a shallow, broad one, and eventually nearly erases it, tracking the black-hole-to-wormhole transition; increasing $\alpha$ at fixed $l=4$ (lower panel) instead deepens and widens the minimum at all radii, reflecting the extra effective attraction from the vector-field charge $Q=\sqrt{\alpha G_N}\,M$, and pushes the ISCO outward. Both panels converge to the same asymptotic value at large $r/M$, confirming asymptotic flatness.

\subsection{Electromagnetic Flux}

\begin{figure*}[ht!]
    \centering
\includegraphics[width=0.49\linewidth]{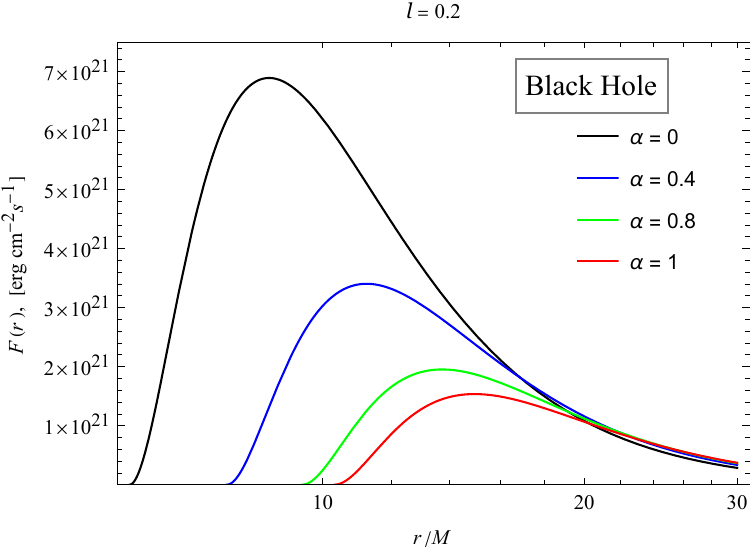}
 \includegraphics[width=0.49\linewidth]{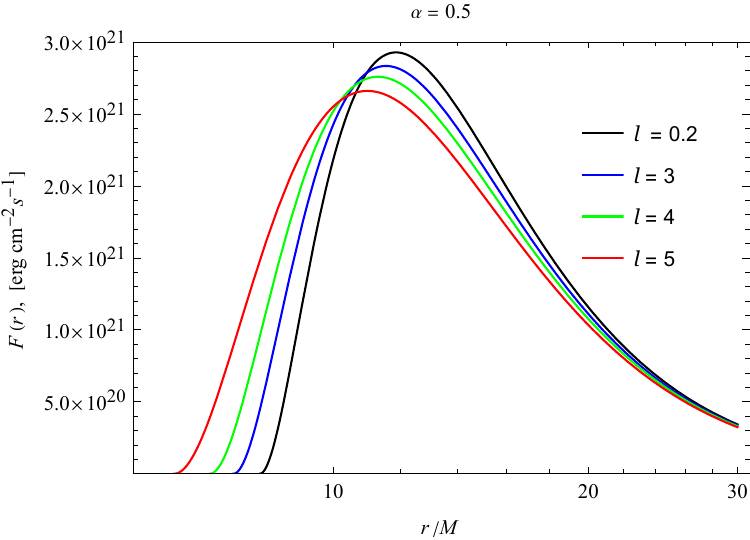}
\caption{Electromagnetic flux $\mathcal{F}(r)$ radiated by a geometrically thin accretion disk around the SV-MOG black hole (left panel) and wormhole (right panel), plotted as a function of the radial coordinate $r/M$ for representative values of the MOG parameter $\alpha$ and the black-bounce parameter $l$.}
    \label{fig:flux}
\end{figure*}

The stress-energy tensor of the accreting fluid is decomposed as~\cite{2022PhRvD.105b3021B}
\begin{equation}
    T^{\mu\nu} = \rho_0 u^\mu u^\nu + 2u^{(\mu}q^{\nu)} + t^{\mu\nu},
    \label{eq:Tmunu}
\end{equation}
where $\rho_0$ is the rest-mass density, $u^\mu$ the four-velocity, $q^\mu$ the energy flux vector satisfying $q^\mu u_\mu = 0$, and $t^{\mu\nu}$ the viscous stress tensor. The surface density of the disk is
\begin{equation}
    \Sigma = \int_{-h}^{h}\langle\rho_0\rangle\,dz,
    \label{eq:Sigma}
\end{equation}
where $\langle\cdot\rangle$ denotes a time and azimuthal average over one orbital period. The vertically integrated torque is
\begin{equation}
    W^r{}_\phi = \int_{-h}^{h}\langle t^r{}_\phi\rangle\,dz.
    \label{eq:torque}
\end{equation}

The conservation of rest mass, $\nabla_\mu(\rho_0 u^\mu) = 0$, gives the time-averaged accretion rate, independent of $r$,
\begin{equation}
    \dot{M}_0 \equiv -2\pi\sqrt{-g}\,\Sigma\,u^r = \text{const}.
    \label{eq:Mdot}
\end{equation}
The conservation of energy, $\nabla_\mu E^\mu = 0$ with $E^\mu = -T^{\mu\nu}(\partial/\partial t)_\nu$, yields
\begin{equation}
    \left[\dot{M}_0 E - 2\pi\sqrt{-g}\,\Omega\,W^r{}_\phi\right]_{,r}
    = 4\pi\sqrt{-g}\,\mathcal{F}\,E,
    \label{eq:energy_conservation}
\end{equation}
and the conservation of angular momentum, $\nabla_\mu J^\mu = 0$ with $J^\mu = T^{\mu\nu}(\partial/\partial\phi)_\nu$, gives
\begin{equation}
    \left[\dot{M}_0 L - 2\pi\sqrt{-g}\,W^r{}_\phi\right]_{,r}
    = 4\pi\sqrt{-g}\,\mathcal{F}\,L.
    \label{eq:angular_momentum_conservation}
\end{equation}
Here $\mathcal{F}(r) \equiv \langle q^z\rangle$ is the time- and azimuth-averaged radiation flux emerging from the disk surface. Eliminating $W^r{}_\phi$ between equations~\eqref{eq:energy_conservation} and~\eqref{eq:angular_momentum_conservation} and using the relation $E_{,r} = \Omega\,L_{,r}$ (which holds on circular geodesics), one obtains the fundamental formula for the radiated flux
\begin{equation}
    \mathcal{F}(r) = -\frac{\dot{M}_0\,\Omega_{,r}}
    {4\pi\sqrt{-g}\,(E-\Omega L)^2}
    \int_{r_{\rm ISCO}}^{r}(E - \Omega L)\,L_{,r}\,dr,
    \label{eq:flux}
\end{equation}
where $\sqrt{-g}\big|_{\theta=\pi/2} = r^2+l^2$ for the metric~\eqref{eq:SV_MOG_metric}, and the lower limit of integration is the ISCO radius at which the torque vanishes by the zero-stress boundary condition, $W^r{}_\phi\big|_{r_{\rm ISCO}} = 0$. Setting 
$l = 0$ and $\alpha = 0$ reproduces the standard Page--Thorne flux formula for the Schwarzschild black hole.

Figure~\ref{fig:flux} shows the disk flux $\mathcal{F}(r/M)$. At fixed $l=0.2$ (left panel), increasing $\alpha$ raises the peak flux and shifts it outward -- from a low, broad profile at $\alpha=0$ to a peak of $\approx7\times10^{21}~\text{erg\,cm}^{-2}\text{s}^{-1}$ at $\alpha=1$, nearly sevenfold higher -- consistent with the outward-shifted ISCO steepening the potential gradient and increasing the dissipation rate. At fixed $\alpha=0.5$ (right panel), increasing $l$ instead suppresses and broadens the flux, lowering the peak from $\approx3\times10^{21}~\text{erg\,cm}^{-2}\text{s}^{-1}$ at $l=0.2$, since a larger bounce parameter weakens the curvature gradient near the throat. In both panels, all curves converge at large $r/M$, showing that the parameter dependence is confined to the strong-field region.

\subsection{Temperature Profile}

\begin{figure*}[ht!]
    \centering
   \includegraphics[width=0.49\linewidth]{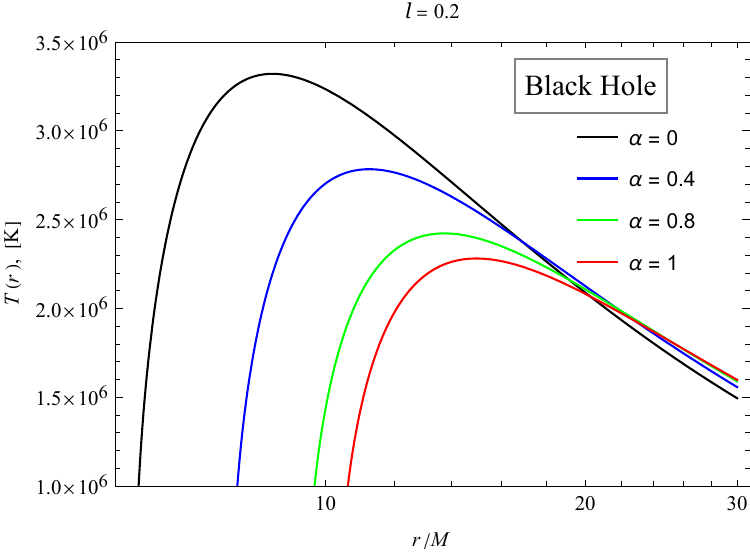}
\includegraphics[width=0.49\linewidth]{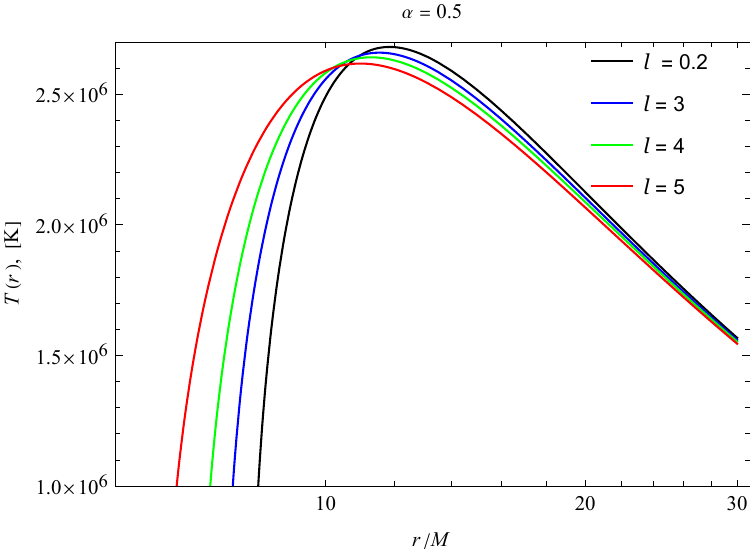}
\caption{Effective temperature $T(r)$ of the accretion disk around the SV-MOG black hole (left panel) and wormhole (right panel) as a function of the radial coordinate $r/M$, shown for several values of the MOG parameter $\alpha$ and the black-bounce parameter $l$. The temperature is expressed in Kelvin (K).}
    \label{fig:temp}
\end{figure*}

Since the disk is assumed to radiate as a perfect blackbody, the Stefan-Boltzmann law relates the local flux to the effective surface temperature. The total power emitted per unit area by a blackbody at temperature $T$ is $\sigma_{\rm SB}T^4$, where $\sigma_{\rm SB} = 2\pi^5 k_B^4/(15 h^3 c^2)$ is the Stefan-Boltzmann constant. Equating this to the disk flux gives
\begin{equation}
    \mathcal{F}(r) = \sigma_{\rm SB}\,T^4(r),
    \label{eq:StefanBoltzmann}
\end{equation}
so the effective disk temperature at radius $r$ is
\begin{equation}
{
    T(r) = \left(\frac{\mathcal{F}(r)}{\sigma_{\rm SB}}\right)^{1/4}.
    }
    \label{eq:temp}
\end{equation}
Since $T \propto \mathcal{F}^{1/4}$, the temperature profile is a smoothed and compressed version of the flux profile. The peak temperature $T_{\rm max}$ and the radius $r_{\rm max}$ at which it is attained depend sensitively on $\alpha$ and $l$:
Figure~\ref{fig:temp} shows the disk effective temperature $T(r/M)$, in Kelvin. At fixed $l=0.2$ (left panel), increasing $\alpha$ \emph{lowers} the peak temperature -- from $\approx3.5\times10^6$~K at $\alpha=0$ to $\approx2.0\times10^6$~K at $\alpha=1$ -- while shifting it outward and broadening the profile; this reflects the same outward ISCO shift as in Fig.~\ref{fig:flux}, which redistributes the dissipated power over a larger disk area even as the integrated flux increases. At fixed $\alpha=0.5$ (right panel), increasing $l$ shifts the peak (of comparable height, $\approx2.5\times10^6$~K at $l=0.2$) outward with only a weak effect on its magnitude, so $l$ influences the profile more weakly than $\alpha$ does. Both panels converge to the same asymptotic temperature, $\approx1.0\times10^6$~K, at large $r/M$.

\begin{figure*}[ht!]
 \includegraphics[width=0.49\linewidth]{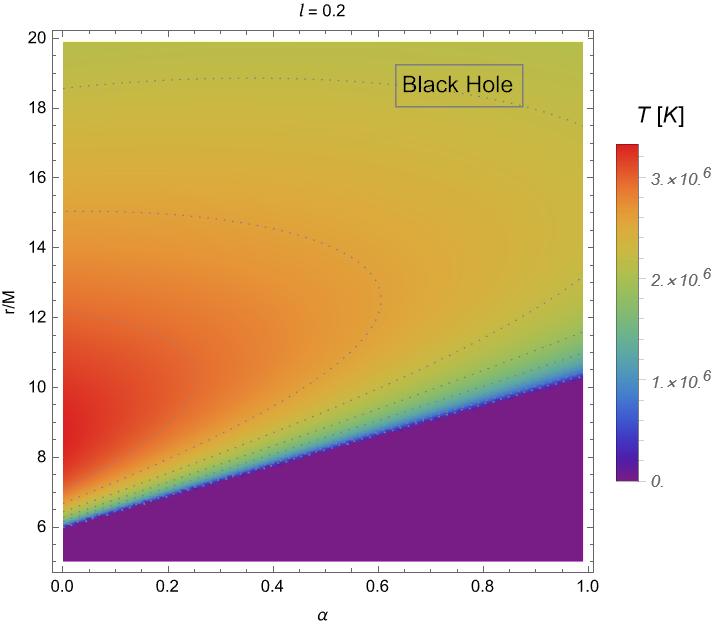}
 \includegraphics[width=0.49\linewidth]{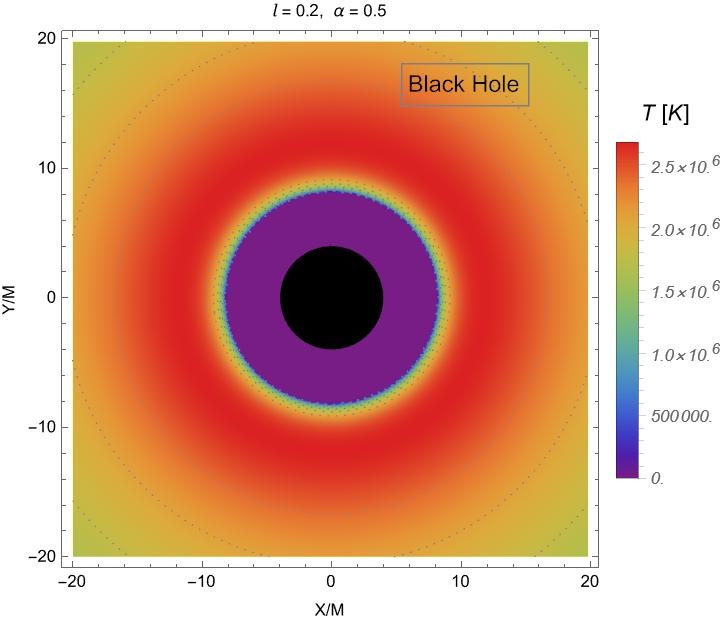}
 \includegraphics[width=0.49\linewidth]{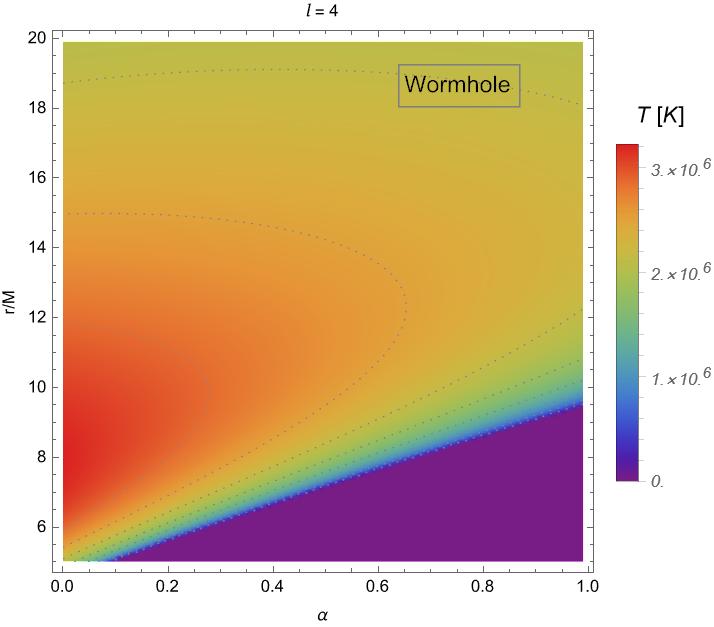}
 \includegraphics[width=0.49\linewidth]{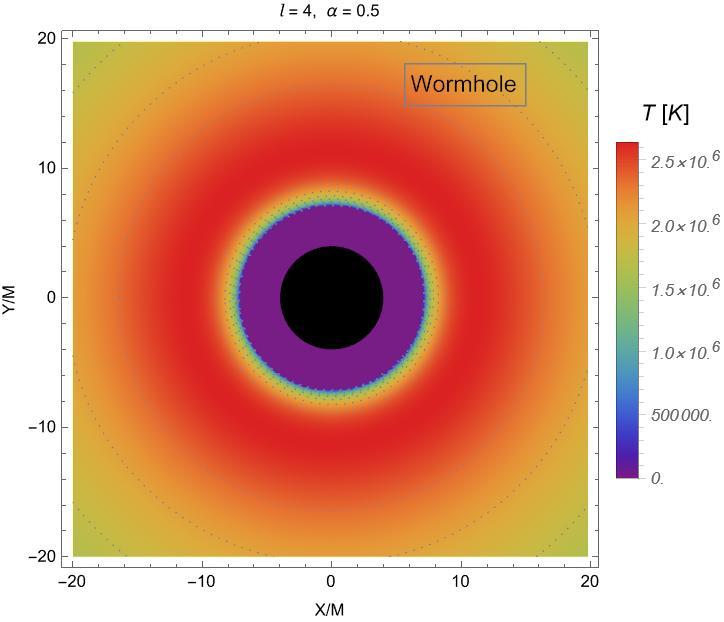}
\caption{Two-dimensional temperature maps of the accretion disk for the SV-MOG black hole (upper panels) and wormhole (lower panels).  Left column: meridional temperature profile $T(r)$ as a function of the radial coordinate for varying MOG parameter $\alpha$ at fixed black-bounce parameter $l$. Right column: false-color map of $T$ on the equatorial $X$--$Y$ plane, where the Cartesian coordinates are defined by $X = r\cos\phi$ and $Y = r\sin\phi$. The color scale encodes temperature in Kelvin (K); brighter regions indicate hotter, inner-disk emission.}
    \label{fig:temperature3}
\end{figure*}

In Fig.~\ref{fig:temperature3}, we present two-dimensional temperature maps of the accretion disk for both the SV-MOG black hole and wormhole configurations. The left column displays the meridional temperature profile $T(r)$ as a function of the radial coordinate $r/M$ and the MOG parameter $\alpha$, while the right column shows the false-color map of $T$ on the equatorial $X$--$Y$ plane, where the Cartesian coordinates are defined by $X = r\cos\phi$ and $Y = r\sin\phi$. The color scale encodes temperature in Kelvin (K), with brighter red regions indicating hotter inner-disk emission and darker purple regions indicating cooler outer-disk emission.
The upper row (black hole, $l=0.2$) shows the disk temperature increasing monotonically with $\alpha$ at fixed $r/M$, hottest near the ISCO (upper-left); the corresponding false-color map at $l=0.2,\ \alpha=0.5$ (upper-right) shows a bright annular hot zone, peaking at $\approx2.5\times10^6$~K just outside the ISCO, surrounding the central shadow, and decreasing smoothly outward with the expected axial symmetry. The lower row (wormhole, $l=4$) is qualitatively similar in the $\alpha$--$r$ plane (lower-left), but the hot region extends further inward, to the throat itself, since no horizon truncates the disk; correspondingly, the false-color map at $l=4,\ \alpha=0.5$ (lower-right) shows a broader hot annulus of comparable peak temperature ($\approx2.5\times10^6$~K) with a smaller, smoothly-bounded central dark region -- reflecting the throat's finite areal radius $l$ in place of a true horizon. Both branches converge to the same cool outer-disk behavior at large radii, confirming that the black-hole/wormhole distinction is confined to the strong-field region.

\subsection{Differential Luminosity}

\begin{figure*}[ht!]
    \centering
    \includegraphics[width=0.49\linewidth]{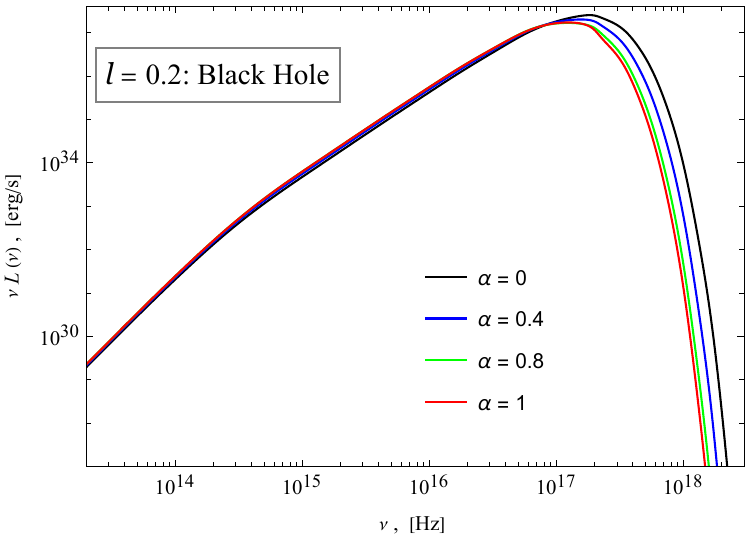}
    \includegraphics[width=0.49\linewidth]{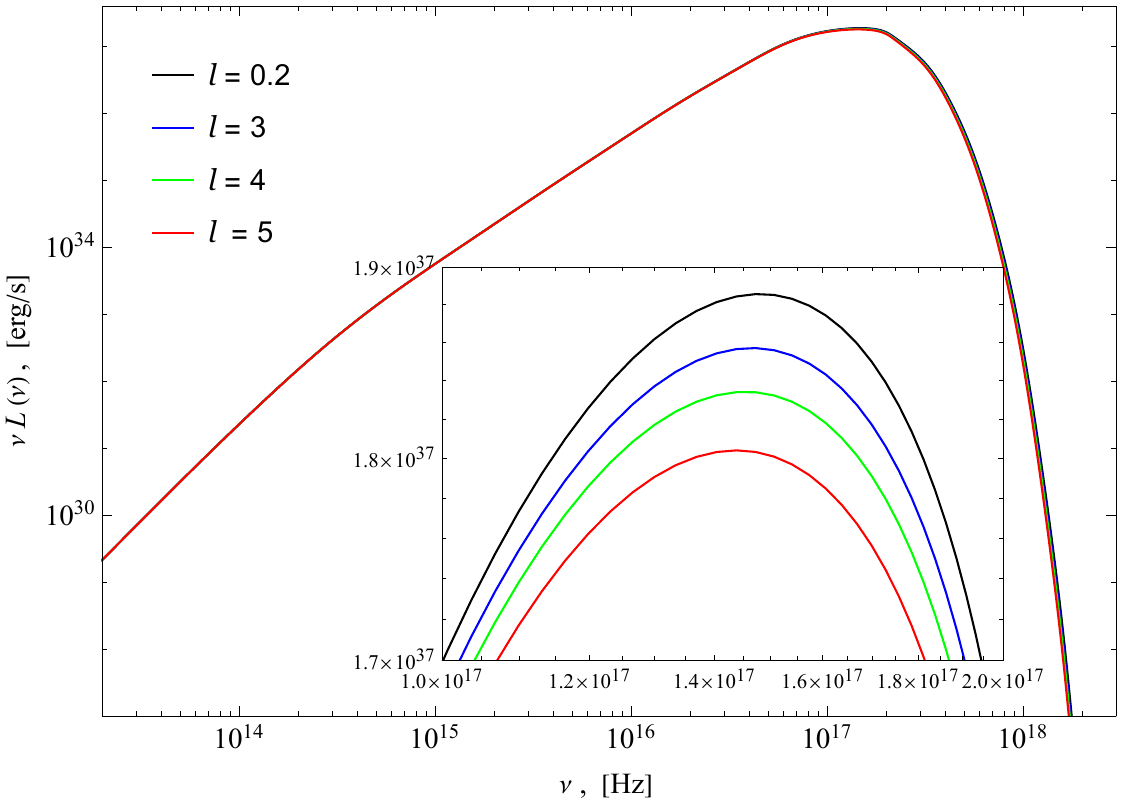}
\caption{Luminosity of the accretion disk surrounding the SV-MOG black hole (left panel) and wormhole (right panel) as a function of the frequency, for various values of the MOG coupling $\alpha$ and the black-bounce parameter $l$. }
    \label{fig:luminosity}
\end{figure*}

The differential luminosity measures the power radiated per logarithmic radial interval and is defined as
\begin{equation}
    \frac{dL}{d\ln r} = 4\pi r\sqrt{-g}\,\mathcal{F}(r)\,E(r),
    \label{eq:dLdlnr}
\end{equation}
where the factor $E(r)$ accounts for the gravitational binding energy of the orbiting fluid. Integrating equation~\eqref{eq:dLdlnr} over all radii gives the total bolometric luminosity
\begin{equation}
    L_{\rm bol} = \int_{r_{\rm ISCO}}^{\infty}
    4\pi r\sqrt{-g}\,\mathcal{F}(r)\,E(r)\,d\ln r
    = \dot{M}_0\,(1 - E_{\rm ISCO}),
    \label{eq:Lbol}
\end{equation}
which defines the radiative efficiency~\cite{2025PhRvD.112d4068J,KHOLMUMINOV2025684,9f1d-7wfh}
\begin{equation}
    \epsilon \equiv 1 - E_{\rm ISCO},
    \label{eq:efficiency}
\end{equation}
where $E_{\rm ISCO}$ is the specific energy of a test particle at the ISCO. For the Schwarzschild geometry, $\epsilon \approx 5.72\%$, while the SV-MOG geometry yields efficiency values that deviate from this due to the shift in the ISCO.

The observed spectral luminosity of the disk, accounting for gravitational and Doppler redshifts, is obtained by integrating the Planck function over the disk area~\cite{Torres2002,doi:10.1142/S0219887826501057,universe10120454},
\begin{eqnarray}\label{eq:luminosity}
    L(\nu) &=& 4\pi d^2 I(\nu)\\ \nonumber
   &=& \frac{8\pi h\cos\gamma}{c^2}
      \int_{r_{\rm ISCO}}^{\infty}\!\int_{0}^{2\pi}
      \frac{\nu_e^3\,r}
           {e^{\,h\nu_e/(k_BT)}-1}\,d\phi\,dr,
\end{eqnarray}
where $d$ is the observer distance, $h$ is Planck's constant, $k_B$ is Boltzmann's constant, $\gamma$ is the disk inclination (taken as $\gamma = 0$ for a face-on configuration), and we neglect photon bending effects~\cite{Novikov1973,2026EPJC...86...58E}. The emitted frequency $\nu_e = \nu(1+z)$ is related to the observed frequency $\nu$ through the total redshift factor, which for a circular equatorial emitter is
\begin{equation}
    1+z = \frac{1 + \Omega\,r\sin\phi\sin\gamma}
               {\sqrt{-g_{tt} - \Omega^2\,g_{\phi\phi}}}.
    \label{eq:redshift}
\end{equation}
The denominator of equation~\eqref{eq:redshift} incorporates both the gravitational redshift (through $g_{tt}$) and the transverse Doppler effect (through $\Omega^2 g_{\phi\phi}$), while the numerator encodes the longitudinal Doppler shift due to the line-of-sight component of the orbital velocity. Setting $\gamma = 0$ eliminates the azimuthal dependence, giving
\begin{eqnarray}\nonumber
    1+z\big|_{\gamma=0}
    &=& \frac{1}{\sqrt{-g_{tt}-\Omega^2\,g_{\phi\phi}}}\\     \label{eq:redshift_faceon}
    &=& \frac{1}{\sqrt{f(r) - \Omega^2(r^2+l^2)}},
\end{eqnarray}
which reduces to the standard Schwarzschild result $(1+z)^{-1} = \sqrt{1-3M/r}$ at the ISCO when $\alpha = 0$, $l = 0$.

Figure~\ref{fig:luminosity} shows $\nu L_\nu$ on a log-log scale. At fixed $l=0.2$ (left panel), all four $\alpha$ curves overlap closely through the rise and the broad maximum near $\nu\sim10^{17}$~Hz; they separate only in the high-frequency tail, where the cutoff frequency \emph{decreases} with increasing $\alpha$ -- i.e. stronger MOG coupling softens the high-frequency emission rather than hardening it, consistent with the redistribution of dissipated power over the enlarged ISCO and with the lower peak disk temperature found at larger $\alpha$ in Fig.~\ref{fig:temp}. At fixed $\alpha=0.5$ (right panel), increasing $l$ likewise suppresses the peak luminosity and softens the spectrum, from $\approx1.9\times10^{34}$ to $\approx1.7\times10^{34}~\text{erg\,s}^{-1}$ between $l=0.2$ and $l=5$, since a larger bounce parameter lowers the disk temperature near the throat; both panels converge at low frequencies, where the emission is insensitive to either parameter.

\begin{figure*}[ht!]
    \centering
    \includegraphics[width=0.49\linewidth]{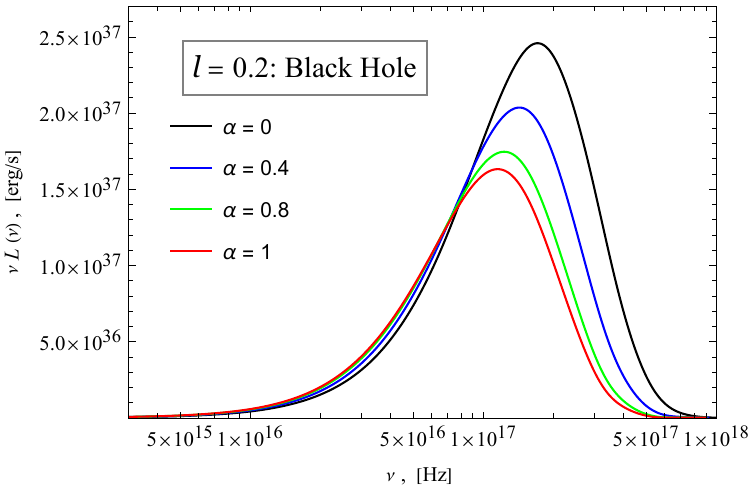}
    \includegraphics[width=0.49\linewidth]{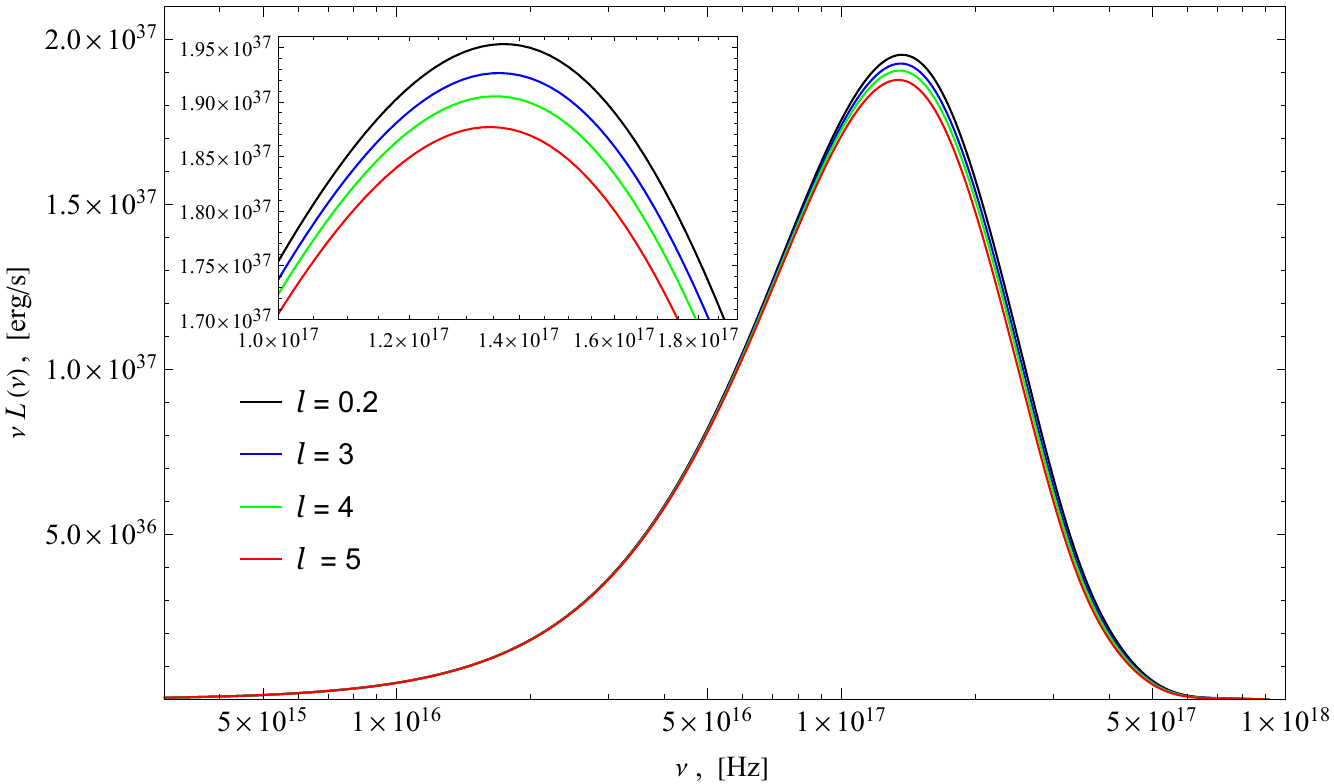}
\caption{Spectral luminosity $\nu L_\nu$ of the SV-MOG accretion disk for the black hole case (left panel) and the wormhole case (right panel), shown as a function of frequency for selected values of $\alpha$ and $l$.}
    \label{fig:luminosity2}
\end{figure*}

Figure~\ref{fig:luminosity2} shows the same quantity on a linear scale, which resolves the peak more clearly. At fixed $l=0.2$ (left panel), increasing $\alpha$ from $0$ to $1$ lowers the bell-shaped peak from $\approx2.5\times10^{37}$ to $\approx1.0\times10^{37}~\text{erg\,s}^{-1}$ and shifts it to lower frequency, in agreement with the softening trend identified in Fig.~\ref{fig:luminosity}. At fixed $\alpha$ (right panel, wormhole branch), increasing $l$ from $0.2$ to $5$ likewise lowers the peak, from $\approx1.95\times10^{27}$ to $\approx1.70\times10^{27}~\text{erg\,s}^{-1}$ -- several orders of magnitude below the black-hole case, reflecting the weaker dissipation near a horizonless throat -- and shifts it redward. Both panels converge at low frequencies, confirming that the parameter dependence is confined to the peak and high-frequency regions of the spectrum.

\section{Conclusion}\label{conclusion}

In this work, we have presented a systematic study of the corrected thermodynamics and accretion-disk radiation of the SV-MOG compact object, obtained by applying the Simpson-Visser regularisation to the Schwarzschild-MOG black hole. The resulting spacetime, parameterized by the MOG coupling constant $\alpha$ and the black-bounce parameter $l$, smoothly interpolates between a regular black hole, a one-way wormhole, and a traversable wormhole, depending on the ratio $l/l_{\rm cr}(\alpha)$.

In the thermodynamic sector, we derived the genuine Hawking temperature and heat capacity for the black-hole branch of the SV-MOG spacetime, and, separately, a physically well-defined effective temperature and heat capacity for the horizonless wormhole branch, constructed from the Lyapunov exponent of the wormhole's unstable photon sphere (Sec.~\ref{thermo}). For the black hole branch, the Hawking temperature exhibits a non-monotonic behavior as a function of the outer horizon radius $r_+$: it rises from zero at the extremal radius, attains a maximum at a critical radius $r_+^*$, and then decreases toward zero at large $r_+$. The divergence of the heat capacity at $r_+^*$ signals a second-order phase transition between thermally unstable and stable phases. We found that increasing $\alpha$ shifts the critical radius to larger values and widens the stable branch, indicating that stronger MOG coupling enhances thermodynamic stability, while increasing $l$ at fixed $\alpha$ narrows the stable window. In the wormhole branch, no event horizon exists, so the standard causal (particle-emission) interpretation of Hawking thermodynamics does not apply; we instead find that the photon-sphere Lyapunov temperature $T_{\rm Lyap}(l)$ decreases monotonically to zero as the light ring is approached from below and eventually disappears, with an associated heat capacity $C_{\rm Lyap}$ that remains negative throughout its window of existence, showing no analogue of the black hole's second-order phase transition and no artificial matching to $T_H,C_V$ at $l=l_{\rm cr}$, since the two constructions encode physically distinct processes. Independently of this thermodynamic characterization, a linear scalar-perturbation analysis (Sec.~\ref{subsec:perturbations}) shows that the effective radial potential remains non-negative throughout both branches for the parameter ranges considered, indicating no evidence of instability against this restricted, monopole-scalar perturbation channel. We stress that this single-channel, linearized, test-field analysis provides a necessary but not sufficient condition for stability; a general dynamical-stability statement would require higher multipoles and coupled gravitational/electromagnetic perturbations, which are left for future work.

We incorporated quantum-gravitational corrections to the Bekenstein-Hawking entropy via logarithmic terms parameterized by the coefficient $\beta$. The corrected entropy $S_c$ reduces to the standard area law at large horizon radii, while deviations become significant as $r_+ \to 0$. A key advantage of the Simpson--Visser regularisation is that the non-zero bounce parameter $l$ prevents the horizon area from vanishing, ensuring that $S_c$ remains bounded from below for all configurations. Increasing $\alpha$ raises the corrected entropy at fixed $r_+$, while increasing $l$ shifts the minimum entropy upward and modifies the divergence structure at intermediate radii. The correction coefficient $\beta$ controls the magnitude of the quantum deviation from the semiclassical value, with larger $\beta$ amplifying the logarithmic correction at small horizon sizes.

In the radiation sector, we computed the electromagnetic flux, effective disk temperature, and spectral luminosity of geometrically thin accretion disks surrounding both black-hole and wormhole configurations within the Novikov--Thorne framework. For the black hole branch, increasing $\alpha$ at fixed $l$ enlarges the ISCO and enhances the peak electromagnetic flux, but it lowers the peak disk temperature and shifts the spectral luminosity peak toward lower frequencies, because the larger dissipated power is spread over a correspondingly larger disk area rather than concentrated close to a small ISCO -- a cooler, spectrally softer, yet more luminous inner disk. Conversely, increasing $l$ at fixed $\alpha$ suppresses the peak flux, has only a weak effect on the peak temperature, broadens the spectral profile, and shifts the luminosity peak toward lower frequencies, reflecting the weakening of the effective curvature gradient near the throat. For the wormhole branch, the disk extends inward to the throat at $r = 0$ with areal radius $l$, generating a distinctive bright central emission region that is absent in standard black hole geometries and could serve as an observational discriminator between the two configurations. The wormhole spectral luminosity is several orders of magnitude lower than the black hole case, reflecting the weaker energy dissipation in the absence of an event horizon.

Taken together, our results demonstrate that the SV-MOG compact object exhibits a rich thermodynamic and radiative phenomenology that is sensitive to both the MOG coupling $\alpha$ and the black-bounce parameter $l$. The observational signatures identified here, in particular, the shift in the spectral peak frequency, the modification of the disk temperature profile, and the distinctive central emission of the wormhole branch, could in principle be constrained by X-ray continuum-fitting observations with current missions such as NICER and future next-generation instruments, providing an independent probe of both modified gravity and regular black hole physics. We emphasize that these are qualitative, model-level predictions of an idealized Novikov--Thorne thin-disk framework (static, non-rotating background; face-on, zero-inclination, geometrically thin and optically thick disk; local blackbody emission; no relativistic ray-tracing or full radiative transfer), and a quantitative comparison with real X-ray spectra would require relaxing these simplifying assumptions. A natural extension of this work would be to incorporate the effects of black hole spin, non-equatorial disk structure, and radiative transfer in the strong-gravity regime, which we leave for future investigation. In particular, building on the rotating SV-MOG constructions of Refs.~\cite{2026ChJPh.102..711K,Khan2026RotSVQPO}, a rotating generalisation of the present analysis would allow the corrected thermodynamics and the accretion-disk observables derived here to be confronted directly with spin-dependent observables such as quasi-periodic oscillation frequencies and shadow asymmetry.

\bibliographystyle{apsrev4-2}
\bibliography{References.bib}
\end{document}